\documentstyle[12pt,aaspp4]{article}
\def\tempest%
{\begin{array}{ccc}
1 & 1 & 1 \\
1 & 1 & 1 \\
4 & 3 & 8
\end{array}}

\def\day{{\rm day}}
\def\min{{\rm min}}
\def\lim{{\rm lim}}
\def\dls{{D_{\rm LS}}}
\def\dos{{D_{\rm S}}}
\def\dol{{D_{\rm L}}}
\def\te{{t_{\rm E}}}

\def\b{{\rm base}}

\def\bu{{\bf u}}
\def\cc{{\rm cc}}
\def\bmu{\hbox{$\mu\hskip-7.5pt\mu$}} 

\begin{document}

\title{A Complete Set of 
Solutions For Caustic-Crossing Binary Microlensing Events}

\author{
M. D. Albrow\altaffilmark{1}, 
J.-P. Beaulieu\altaffilmark{2},
J. A. R. Caldwell\altaffilmark{3}, 
D. L. DePoy\altaffilmark{4}, 
M. Dominik\altaffilmark{2}, 
B. S. Gaudi\altaffilmark{4}, 
A. Gould\altaffilmark{4}, 
J. Greenhill\altaffilmark{5}, 
K. Hill\altaffilmark{5},\\
S. Kane\altaffilmark{5,6}, 
R. Martin\altaffilmark{7},  
J. Menzies\altaffilmark{3}, 
R. M. Naber\altaffilmark{2}, 
R. W. Pogge\altaffilmark{4},
K. R. Pollard\altaffilmark{1},\\ 
P. D. Sackett\altaffilmark{2}, 
K. C. Sahu\altaffilmark{6}, 
P. Vermaak\altaffilmark{3},  
R. Watson\altaffilmark{5}, 
A. Williams\altaffilmark{7}
}
\author{The PLANET Collaboration}

\altaffiltext{1}{Univ. of Canterbury, Dept. of Physics \& Astronomy, 
Private Bag 4800, Christchurch, New Zealand}
\altaffiltext{2}{Kapteyn Astronomical Institute, Postbus 800, 
9700 AV Groningen, The Netherlands}
\altaffiltext{3}{South African Astronomical Observatory, P.O. Box 9, 
Observatory 7935, South Africa}
\altaffiltext{4}{Ohio State University, Department of Astronomy, Columbus, 
OH 43210, U.S.A.}
\altaffiltext{5}{Univ. of Tasmania, Physics Dept., G.P.O. 252C, 
Hobart, Tasmania~~7001, Australia}
\altaffiltext{6}{Space Telescope Science Institute, 3700 San Martin Drive, 
Baltimore, MD. 21218~~U.S.A.}
\altaffiltext{7}{Perth Observatory, Walnut Road, Bickley, Perth~~6076, Australia}

\begin{abstract}

	We present a method to analyze binary-lens microlensing light curves
with one well-sampled fold caustic crossing.  In general, 
the surface of $\chi^2$ shows extremely complicated behavior over the 9-parameter space that 
characterizes binary lenses.  This makes it difficult to systematically
search the space and verify that a given local minimum is a global
minimum.  We show that for events with well-monitored caustics,
the caustic-crossing region can be isolated from the rest of the light curve 
and easily fit to a 5-parameter function.  Four of these caustic-crossing
parameters can then be used to constrain the search in the larger 9-parameter space.  
This allows a systematic search for all solutions and thus identification of all local minima.  
We illustrate this technique using the PLANET data for MACHO 98-SMC-1, 
an excellent and publicly available caustic-crossing data set.  We show 
that a very broad range of parameter 
combinations are compatible with the PLANET data set, demonstrating that
observations of binary-lens lightcurves with sampling of only one caustic
crossing do not yield unique solutions.  The corollary to this is that
the time of the second caustic crossing cannot be reliably predicted on the basis
of early data including the first caustic crossing alone.  We investigate 
the requirements for
determination of a unique solution and find that occasional observations
of the first caustic crossing may be sufficient to derive a complete solution.
 
\end{abstract}


\section{Introduction}
	Binary-lens microlensing events, especially those with
caustic crossings, have a number of potentially important applications.  
First, if the caustic crossing is well sampled, 
the proper motion of the lens
relative to the observer-source line of sight can be measured.  Since different populations
have different proper-motion distributions, such a measurement can help
determine the nature of the lens.  For example, five groups observed the
event MACHO 98-SMC-1 found by the MACHO collaboration (Alcock et al.\ 1999)
in observations toward the Small Magellanic Cloud (SMC) and all concluded
that its proper motion is consistent with the lens being in the SMC rather
than the Galactic halo (Afonso et al.\ 1998 [EROS]; 
Albrow et al.\ 1999a [PLANET]; Alcock et al.\ 1999 [MACHO]; 
Udalski et al.\ 1998 [OGLE]; Rhie et al.\ 1999 [MPS]).  This provides
an important clue regarding the controversy (e.g.\ Sahu 1994; Gould 1995) 
over the location and nature of the lenses currently being discovered toward
the Magellanic Clouds (Aubourg et al.\ 1993; Alcock et al.\ 1997a).
Second, caustic-crossing binaries are one of the few classes of microlensing events for
which it is possible, at least in principle, to obtain a complete solution
of the mass, distance, and velocity of the lens (Hardy \& Walker 1995;
Gould \& Andronov 1999).  Third, caustic-crossing events (both binary and points lens)
can be used to measure the limb-darkened profile of the source star (Albrow et al.\ 1999b).
Fourth, binary-lens events can potentially tell
us about the distributions of binary mass ratios and separations.  The
light-curve solution directly yields the mass ratio and also gives the
projected separation in units of the Einstein ring.  By
calibrating the binary-lens detection efficiency (Gaudi \& Sackett 1999),
the observed distribution can be compared with that predicted for
various models.  Finally, planet-star systems are, from a microlensing
standpoint, extreme mass-ratio binaries and hence can be discovered by
looking for binary-lens type events (Mao \& Paczy\'nski 1991).

	For most of these applications, one must correctly and uniquely
measure the parameters that describe the observed binary lens
and quantify the uncertainties in this solution.  Or,
if an unambiguous determination is not possible, one must at least find
the entire set of degenerate solutions. 

	Nine parameters are required to specify the most basic caustic-crossing
binary-lens event.  These are usually taken to be $t_0$, $u_0$, $\te$, $q$, $d$, 
$\alpha$,
$\rho_*$, $F_s$, and $F_b$.  Here $t_0$ is the time of closest approach to
the origin of the binary,
$\te$ is the Einstein crossing time, and $u_0$ is the angular impact 
parameter at time $t_0$ in units of the angular Einstein radius, $\theta_{\rm E}$,
\begin{equation}
\theta_{\rm E} = \biggl({4 G M \dls \over \dol\dos c^2}\biggr)^{1/2},
\label{eqn:thetae}
\end{equation}
 $\dol$, $\dos$, and $\dls$ are the observer-lens,
observer-source, and lens-source distances, 
and the mass $M$ is the 
total mass of the binary.
Note that $\theta_{\rm E}=\mu\te$, where $\mu$ is the
proper motion.  The three parameters specific to the binary character of the 
lens are the mass ratio $q=M_2/M_1$ of the secondary to the primary 
($0<q\leq 1$),
the projected angular binary separation $d$ in units of $\theta_{\rm E}$,
and the angle $\alpha$ $(0\leq\alpha<2\pi)$
between the binary-separation vector ($M_2$ to $M_1$)
and the proper motion of the source relative to the origin of the binary.   
Our convention is that the center of the binary lies on the right hand
side of the moving source, and we adopt the midpoint of the lenses as the origin of
the binary.
The angular size of the source in units of $\theta_{\rm E}$ is $\rho_*$,
the source flux is $F_s$, and $F_b$ is the light from any unlensed sources
(including the lens) that enters the aperture.  If the event is observed from
more than one observatory, then two additional parameters are required for
each additional observatory 
to account for the different fluxes and backgrounds registered by
different telescopes.  One may include more than these basic parameters to account
for other higher-order effects, such as limb darkening of the source, orbital motion
of the binary, and parallax effects due to the motion of the Earth, but we will
ignore these effects in this paper.  This means that we will be implicitly assuming
that the source can be approximated as uniform.

	Using the above parameterization, fitting binary-lens light
curves poses a
significant challenge for several reasons.   First, $\chi^2$ is very sensitive to small
changes in most of the parameters, and furthermore responds in a complicated
manner.  The sheer size of parameter space combined with the sensitivity
of $\chi^2$ to subtle changes in the parameters make 
brute force searches practically
impossible.  Second, choosing suitable 
initial guesses for possible solutions is difficult
because most of the parameters have no direct relationship to observable features in
the light
curve.  Thus, even if one finds a trial solution, it is difficult to be sure
that all possible solutions have been found.  Finally, the magnification of 
a binary lens
is nonanalytic.  While this poses no significant challenge for calculating light curves for 
events that can be approximated as having a point 
source, such as binary-lens events
with no caustic crossings, finite-source caustic crossing light curves
are notoriously difficult to calculate.  Although many efficient and robust methods have been
proposed to do this (Kayser \& Schramm 1988; Gould \& Gaucherel 1997; 
Wambsganss 1997; Dominik 1998), 
they are invariably time consuming.
This is a serious detriment to fitting a light
curve because of the large number of models that
must be calculated.  

Mao \& Di Stefano (1995) attacked the problem of fitting binary-lens light
curves
by developing a densely-sampled
library of point-source binary microlensing
events, each of which is characterized by catalogued ``features'' such as the 
number of maxima, heights of peaks, time between peaks, etc.  They can then examine individual 
events, characterize their ``features,'' and search their library for
events that are consistent with these features.  This alleviates many of the problems
discussed above, as it reduces the
search to a relatively few regions of parameter space.  Mao \& Di Stefano
(1995) report that their method is robust for caustic crossing events since
these have well defined features.  However, this method does have some 
shortcomings that make it difficult to apply to well-sampled 
caustic-crossing binary-lens events.  First, the method
relies on the approximate magnification of the observed peaks to reduce the 
possible space of solutions.  However, the magnification of the observed peaks depends
on the baseline magnitude, which can be unknown or poorly determined.  Furthermore, 
even if the baseline is exactly measured, the magnification is not a direct observable,
as it depends not only on
the binary model and trajectory, but also on the amount of 
blended light, $F_b$.  Finally, the peak magnification also depends on the unknown size of the 
source $\rho_*$.    While it may be possible to extend the method of Mao \& Di Stefano (1995)
to take into account these difficulties, the search space would increase by
two dimensions and thus the efficiency would decrease.
Di Stefano \& Perna (1997) suggested that binary lenses could be fitted by decomposing the
observed light 
curve into a linear combination of basis functions.  The coefficients of
these functions could
 then be compared to those fitted to a library of events in order
to isolate viable regions of parameter space.  This is essentially the 
same method as Mao \& Di Stefano (1995), except that, rather than use gross features 
to identify similar light
curves, one uses the coefficients of the polynomial expansion,
which is more quantitative and presumably more robust.  However, this method has the
same shortcomings as that of
Mao \& Di Stefano (1995) for the same reasons.  Also, the method of
Di Stefano \& Perna (1997) requires that, before the basis function fitting, 
one map the observed light
curve onto the same temporal interval for which the event library light
curves were fitted.  This is impossible to do for 
only partially sampled events, or events where the fraction of blended light is unknown.

	Here we propose an althernate method to systematically search for solutions
in the specific case of a binary lens with one well-sampled caustic crossing.
Initially, binary-lens events were monitored only by the primary search
groups and so were observed only once or twice per night.  Since caustic
crossings generally take less than one day, this implied that the crossings
were not well sampled.  For example, the first binary-lens event with caustic
crossings,
OGLE-7, was observed by OGLE only once near the first caustic and not at
all near the second (Udalski et al.\ 1994), although MACHO did serendipitously
observe one point on the second crossing of this event
thereby resolving the source (Mao et al.\ 1994).  Dominik (1999a) showed
that a large variety of binary-lens parameters are consistent with the
photometric data for this event as well as for another caustic-crossing 
binary, DUO-2.

	However, at present the three primary search groups, OGLE, MACHO, and 
EROS, all have alert systems by which they can recognize
microlensing events in real time.  Three other groups, 
GMAN (Alcock et al.\ 1997b), PLANET, and MPS 
then monitor these alerted events much more frequently.
Once a source crosses the first caustic, it is possible to predict the second
crossing at least a day in advance on the basis of these frequent follow-up
observations by observing the rise to the crossing (although it is not possible to
predict the second caustic crossing from observations of the first alone, as we demonstrate
in \S\ 4.2).  The second caustic crossing can then be observed very intensively.  
Indeed, one caustic
crossing was even observed spectroscopically by making use of 
target-of-opportunity time (Lennon et al.\ 1996).  Thus, well-sampled caustic
crossings should become more common in the future.

	We present our method for searching for binary-lens solutions in \S\ 2.
In \S\ 3 and \S\ 4, we illustrate the method using PLANET data for
MACHO 98-SMC-1.  We show that a broad range of parameter combinations are 
consistent with the PLANET data.  In \S\ 5, we therefore examine what sort
of data are required to break these degeneracies.
	
	We emphasize that our treatment of MACHO 98-SMC-1  is not
intended to be definitive, but merely illustrative.  A thorough investigation
of this event will be made by Afonso et al.\ (1999) by combining
data from all five groups.  

\section{The Method}

	We assume that the binary-lens light
curve can be decomposed into two parts.  The first part characterizes the caustic
crossing itself, and is described by a five-parameter semi-analytic function.  The five
parameters are not directly related to any of the traditional parameters,
but are more directly related to 
observables, so that 
$\chi^2$ is less sensitive to small changes in these parameters.  Furthermore,
the function is semi-analytic, and thus very simple and quick to compute.  We fit the
data near the caustic crossing to this function.  Four of the five parameters extracted from the 
fit, along with a measurement of the baseline, are then used to constrain the search
of parameter space.  We then search for fits to the non-caustic crossing 
light-curve data in this restricted space.  We calculate the magnification of these images 
from the full binary-lens equation with the
standard parameters.  Since the magnification arising from the 
diverging images associated with the caustic is not being considered, $\chi^2$ behaves 
much more sensibly.  Furthermore, no finite-source effects need be considered when fitting
to the non-caustic crossing data, 
greatly improving the computational efficiency of the search.  The
end result of this search is a complete set of trial solutions.  We then perform
refined searches begining with these trial solutions, incorporating  
all the data, and using a variant of the method just described.

In the next section, we describe in detail the method of fitting and extracting parameters
from the five-parameter function that describes generic caustic crossings.  The section following
then describes how the parameters extracted from the caustic-crossing fit can be used to constrain
the search for the global fit to the remaining data, and an effective method for performing 
this search. Figure \ref{fig:flow} is a flow chart which illustrates the
relations among the various steps of the method.

\subsection{Parameterized Fit to the Caustic Crossing Data}

Imagine a point at the center of a source as it crosses a caustic.  
While inside the caustic, the point source has
five images.  As it crosses the caustic, the magnifications of two of these 
images diverge toward a square-root singularity, until the images
suddenly disappear. 
If we neglect any changes of the lens properties in the neighborhood
of the caustic crossing, then the magnification of these two divergent
images can be written (e.g.\ Schneider \& Wei\ss 1986a),
\begin{equation}
A_{\rm div}^0(\bu) = \biggl({\Delta u_\perp\over u_r}\biggr)^{-1/2}
\Theta(\Delta u_\perp),
\label{eqn:adiv}
\end{equation}
where
\begin{equation}
\Delta u_\perp \equiv  \Delta {\bf u \cdot n}_\cc, \qquad
\Delta \bu\equiv \bu - \bu_\cc,
\label{eqn:duperpeq}
\end{equation}
$\bu_\cc$ is the position of the caustic crossing, ${\bf n}_\cc$ is the 
unit vector at $\bu_\cc$ pointing inward normal to the caustic,
$\Theta$ is a step function, and
$u_r$ is the characteristic rise length of the caustic.
The other three images are unaffected by the caustic crossing, so their
total magnification can be Taylor expanded, 
\begin{equation}
A_{\rm non-div}^0(\bu) = A_\cc + {\bf Z}\cdot \Delta \bu,
\label{eqn:anondiv}
\end{equation}
where $A_\cc$ is the magnification of the three images at the caustic
crossing, and $\bf Z$ is the gradient of the magnification with respect to
$\bu$.  Hence the full magnification in the neighborhood of the caustic
crossing can be approximated as,
\begin{equation}
A^{0}(\bu) = \biggl({\Delta u_\perp\over u_r}\biggr)^{-1/2}
\Theta(\Delta u_\perp) + A_\cc + {\bf Z}\cdot \Delta \bu.
\label{eqn:a0u}
\end{equation}
For an extended source of angular radius 
$\theta_*\equiv\rho_*\theta_{\rm E}$, the
magnification is given by the convolution of $A^0$ with
the source surface brightness
profile, which yields (e.g.\ Schneider et al.\ 1992, p.\ 215f),
\begin{equation}
A(\bu) = \biggl({u_r\over \rho_*}\biggr)^{1/2}
G\biggl(-{\Delta u_\perp\over \rho_*}\biggr) + A_\cc + {\bf Z}\cdot \Delta \bu.
\label{eqn:abu}
\end{equation}
Note that $\Delta u_\perp$ is positive and the argument of $G$ is
negative when $\bf u$ is inside the caustic.

Here $G$ is a characteristic profile function which depends only on the
shape of the stellar
profile, and not on the size of the source.  That is, the source size
affects the width of the caustic crossing only through the argument 
$\Delta u_{\perp}/\rho_\star$ of $G$ and the magnification only
through the factor $\rho_\star^{-1/2}$.
For uniform surface brightness, the profile function $G$ reads 
(Schneider \& Wei\ss 1986b),
\begin{equation}
G_0(\eta) \equiv 
\frac{2}{\pi}\int_{\rm max(\eta,-1)}^{1}\left(\frac{1-x^2}{x-\eta}
\right)^{1/2}dx \Theta(1-\eta),\label{eqn:atrue}
\end{equation}
which can be expressed in terms of elliptical integrals.
The case of limb-darkened profiles has been discussed 
by Schneider \& Wagoner (1987).
Consider an extended source moving over the caustic with proper motion
$\mu=\theta_E/t_E$, at an angle $\phi$ relative to the caustic.  The time required
for the radius to cross the caustic is 
\begin{equation}
\Delta t = {\theta_* \over \mu \sin\phi} = \rho_*\te\csc\phi\, .
\label{eqn:deltat}
\end{equation}
Note that the width of the caustic crossing $\Delta t$ can be
measured from the caustic-crossing data alone, while the three quantities
whose product forms $\Delta t$ ($\rho_*$, $t_{\rm E}$, and $\csc\phi$) can
only be determined from an analysis of the complete light curve (see
\S\ 2.3). 
The angular separation (normalized to $\theta_{\rm E}$) of the source
from the caustic crossing as a function of
time is,
\begin{equation}
\Delta\bu = {\bmu(t-t_\cc)\over\theta_{\rm E}},
\label{eqn:dbu}
\end{equation}
where $\bmu$ is the vector proper motion, and $t_\cc$ is the time of the
caustic crossing.  This implies,
\begin{equation}
\Delta u_\perp = {\mu(t-t_\cc)\sin\phi\over \theta_E}
= {t-t_\cc\over t_{\rm E}}\sin\phi.
\label{eqn:duperp}
\end{equation}
Hence the magnification as a function of time is given by,
\begin{equation}
A(t) = \biggl({t_{\rm r}\over \Delta t}\biggr)^{1/2} 
G_0\biggl({t-t_{\rm cc}\over \Delta t}\biggr)
+ A_{\rm cc} +\omega(t-t_{\rm cc})\,,
\end{equation}
where 
\begin{equation}
t_r = u_r t_{\rm E}\csc\phi,
\label{eqn:trdef}
\end{equation}
is the characteristic rise time of the caustic crossing, and
$\omega\equiv \bmu\cdot {\bf Z}/\theta_{\rm E}$.

If the flux of the source star is $F_{\rm s}$ and the flux of the
blend is $F_{\rm b}$, the total flux is given by
\begin{equation}
F(t) = F_s A(t) + F_b = Q^{1/2} G_0\biggl({t - t_{\rm cc}\over\Delta t}\biggr)
(\Delta t)^{-1/2}
+ F_{\rm cc} + \tilde \omega(t - t_{\rm cc}),\label{eqn:ftrue}
\end{equation}
where $Q = F_s^2 t_r$, $F_{\rm cc} = F_s A_{\rm cc} + F_b$, and
$\tilde \omega = F_s \omega$. Thus, a caustic crossing
can be fit to a five-parameter 
function of the form of equation (\ref{eqn:ftrue}),
the parameters being $Q, t_{\rm cc}, F_{\rm cc}, \Delta t$, and $\tilde\omega$.
Below, we will use the three parameters $Q$, $t_{\rm cc}$, and 
$F_{cc}$ to constrain the search for fits to the {\it non-caustic-crossing}
points on the light curve.  
The caustic-crossing time scale $\Delta t$ 
summarizes information about the caustic crossing only, and does not
affect the remainder of the light
curve or its analysis.
The slope $\tilde \omega$ was introduced only to allow a more accurate 
estimate of $F_{cc}$ and will be of no further interest.  

It is also possible to parameterize the total flux by,
\begin{equation}
F(t)= F_{\rm cc}\,\left[ \left({t_{\rm r, eff}\over\Delta t}\right)^{1/2}\, G\biggl({t - t_{\rm cc}\over\Delta t}\biggr) + 1 +{\omega_{\rm eff}}(t-t_{\rm cc}) \right]
,\label{eqn:ffalse}
\end{equation}
where 
\begin{equation}
t_{\rm r, eff} = \left( {F_{\rm s} \over F_{\rm cc}}\right)^2\,t_{\rm r}={Q \over F_\cc^2}\,,
\label{eqn:defp}
\end{equation}
is an ``effective'' rise time, and ${\omega_{\rm eff}}={\omega/F_{\rm cc}}$.
This parameterization seems to be more appealing, as it replaces 
the unintuitive parameter $Q$ with $t_{\rm r, eff}$, 
the effective rise time of the caustic crossing.  
Unfortunately, in this parameterization, $F_{\rm cc}$
and $t_{\rm r, eff}$ are very highly correlated: we find below for
a specific example that the fractional error in $t_{\rm r, eff}$
is about 7 times larger than the fractional error in $Q$ which makes
$t_{\rm r, eff}$ substantially less suitable for numerical calculations.
We will therefore use the parameterization in equation (\ref{eqn:ftrue}).

	Note that in the neighborhood of the end of the caustic crossing,
\begin{equation}
G_0(\eta)\rightarrow 2^{1/2}(1-\eta)\Theta(1-\eta), 
\label{eqn:geqlimit}
\end{equation}
and thus
an abrupt change of slope occurs at $\eta=1$.  Hence, while for most points on the
light curve it is appropriate to use simply the midpoint of the exposure for
the time, this approximation breaks down when the time between the midpoint of
the exposure and the end of the caustic crossing
($\eta\sim 1$), $\delta t = t - \Delta t - t_{\rm cc}$,
is less than half the exposure time, $t_{\rm exp}$, i.e. $|\delta t| < t_{\rm exp}/2$.
For this case we integrate equation (\ref{eqn:geqlimit}) over the exposure
time and find,
\begin{equation}
G_0\biggl({t-t_\cc\over \Delta t}\biggr)
\rightarrow 2^{-1/2}{(\delta t -t_{\rm exp}/2)^2\over t_{\rm exp} 
\Delta t}, \qquad \biggl(|\delta t| < {t_{\rm exp}\over 2}\biggr).
\label{eqn:finiteexp}
\end{equation}

\subsection{Relations Between Parameterizations}

	As shown in the previous subsection, the caustic-crossing fit yields 
the four parameters $Q$, $t_{\rm cc}, \Delta t$, and $F_{\rm cc}$.  Three of the remaining parameters 
are the same as in the conventional parameterization:
the Einstein time scale, $\te$, the normalized projected separation
between the lenses, $d$, and the mass ratio, $q$.  
The eighth parameter is the baseline flux $F_\b$ which is often, 
but not always, well measured.
For the final parameter, we adopt the path length $\ell$ 
along the caustic curve(s) for
the configuration ($d,q$).  This is a logical choice, since
we know that the light curve contains a caustic crossing, and the trajectory must therefore
cross a caustic at some value of $\ell$.  Below we show how the local properties of the
binary-lens at $\ell$ can be used to relate our non-standard parameters to the more
familiar parameters.  In our parameterization,
the binary-lens event is described by the 9 parameters $(Q, t_{cc}, \Delta t, F_{cc},
d,q,\ell,\te, F_\b)$ rather than by the 9 ``standard'' parameters $(t_E, t_0, u_0,
d,q,\alpha, \rho_*, F_s, F_b)$.  In order to use the caustic crossing parameters
$(Q, t_{cc}, F_{cc})$ to constrain the fit to the non-caustic crossing data, we must 
know the relation between the two parameter sets.  This is trivial for $\te, d$, and
$q$. Given a binary configuration ($d,q$), one can determine at each $\ell$ 
the following five
local properties of the binary lens.  The first two are simply the x and y positions of
the caustics at $\ell$, $u_{\rm cc, x}(\ell)$ and $u_{\rm cc, y}(\ell)$ 
with respect to the standard coordinate system (i.e. the origin located
at the midpoint of the binary and the x-axis coincident with the binary axis).  These
values can be determined using the algorithm of Witt (1990). The third property is the
angle of the caustic with respect to the binary axis at $\ell$, $\gamma (\ell)$, which can
be found by the same algorithm and by fitting a line to positions
offset by $\delta \ell$ from $\ell$.  The last two properties 
must be calculated by solving the full
binary-lens equation.  The near-caustic magnification $A_{\rm cc}$ is the sum of the 
magnifications of the three non-diverging images at the position of the 
caustic.  The caustic divergence, $u_r$, is defined by equation (\ref{eqn:adiv}) and can be 
determined by fitting an inverse square-root function to the sum of the magnifications
of the two diverging images in the neighborhood of $\ell$. 
Note that all five quantities are functions of
$(\ell,d,q)$.  Using these quantities, the
relations between the standard parameters and those used in this paper are
simple to determine and are given in Table 1.  Figure \ref{fig:onep} shows
the relation between the two sets of parameters graphically for the parameters
that do not involve the finite-size of the source.  Figure \ref{fig:twop} shows
a detailed view of the finite source crossing the caustic.  Note that several
of the quantities shown in Figure \ref{fig:twop} are not discussed in the
text until equation (\ref{eqn:analytica}) in \S\ 4.1, below.

\subsection{Fitting Non-Caustic-Crossing Data: Idealized Case}

	We now use our parameterization and 
the results of the fit to the caustic-crossing data to find
corresponding binary-lens configurations that contain the observed
caustic crossing.  For illustrative purposes, let us initially assume that both the 
baseline flux of the event, $F_\b$, and the three caustic-crossing
parameters $Q$, $F_{\rm cc}$,
and $t_{\rm cc}$ have been measured with high precision.  
(Recall that the fourth caustic-crossing parameter, $\Delta t$, is not used
in the analysis of the non-caustic-crossing data.)\ \ 
The search for solutions
would then be reduced to a four-dimensional space and could be conducted as
follows.  First, one begins with a binary configuration $(d,q)$, varying
the parameter $\ell$ over the total length of the caustic.  At each $\ell$
in geometry $(d,q)$, 
one has two equations relating the source and background fluxes:
$F_\b = F_s + F_b$ and $F_{\rm cc} = F_s A_{\rm cc} + F_b$.  Thus,
\begin{equation}
F_s = {F_{\rm cc} - F_\b\over A_{\rm cc} -1}.\label{eqn:fseval}
\end{equation}
If $F_s>F_\b$ (i.e., $A_{\rm cc}<F_{\rm cc}/F_\b$), then there would be 
negative
background flux.  Hence any position $\ell$ yileding $F_s>F_\b$ does not
correspond to a physical solution, and one can move on to the next value 
of $\ell$.  At each physical $\ell$, $\te$ is varied and for each
$\te$, the angle $\phi$ at which the source crosses the caustic is
determined using equation (\ref{eqn:trdef}) and the definition 
$Q\equiv F_s^2t_r$,
\begin{equation}
\sin\phi = 
{u_r\,\te\,F_s^2\over Q},\label{eqn:sinphieval}.
\end{equation}
Of course, $\phi$ must satisfy $\sin\phi\leq 1$, which means that only
values of $\te$ in the range
\begin{equation}
\te\leq {Q\over u_r\,F_s^2},\label{eqn:telim}
\end{equation}
need to be searched.  Note that $\phi$ is restricted to lie in the range
$0\leq\phi\leq \pi$, and the orientation of $\gamma$ is set to enforce
the relation in Table 1: $\alpha=\phi+\gamma$.
At this point all of the standard parameters needed to
evaluate the magnifications at all the times of the observations have been determined.  
Since $F_s$ and $F_b (=F_\b-F_s)$ are
completely determined for this geometry, these magnifications can be used
to predict the flux, $F(t) = F_s A(t) + F_b$, and these predictions can
be compared to the data using $\chi^2$.  However, before doing the
calculation for the entire (non-caustic-crossing) light curve, the following
checks should be done.  From inspection of the light curve, it is often
clear which measurements are inside the caustic and which are outside the caustic.
One could then evaluate the number of images at the most restrictive of these
measurements (i.e., the last measurement before the first crossing and
 the first measurement
after this crossing), and determine whether the model is consistent with these 
observational constraints.  If it is not, no further 
evaluation need be done and one can continue to the next parameter combination.  Thus one
can calculate $\chi^2$ for each combination $(d,q,\te,\ell)$ and find the best fit
(or fits) to the data.  The search is over a four-dimensional space, but under restricted
circumstances.

\subsection{Non-Caustic-Crossing Data: Realistic Case}

	In practice, $Q$, $F_{\rm cc}$, and
$F_\b$ are not known with infinite precision, and so one must take account
of the uncertainties in these parameters.  For well-sampled caustic crossings,
the time of caustic crossing
$t_{\rm cc}$ is measured to much higher precision than is required, so for this
purpose we assume that it is known perfectly.  The parameter $\Delta t$ has
no effect on the non-caustic-crossing data, so uncertainties in this parameter are
unimportant.  The uncertainties in $Q$, $F_{\rm cc}$, and
$F_\b$ introduce two 
major changes into the above procedure.  First, one must consider a {\it
range} of $\sin\phi$ at each parameter combination $(d,q,\ell,\te)$ rather
than a single value.  That is, there is a fifth dimension to the search,
albeit over a truncated domain.  Second, once a parameter combination 
$(d,q,\ell,\te)$ is chosen, and the range in $\phi$ to be explored 
is determined,
one must {\it fit} for the two flux
parameters $F_s$ and $F_b$ since these are no longer determined with 
infinite precision.  This appears to add two dimensions to the search,
but in fact this is a linear fit and can be computed much more quickly
than the other steps required for each combination $(d,q,\ell,\te,\phi)$
(see also Rhie et al.\ 1999).  Thus,
the search is effectively increased to 4.5 dimensions.  Good
constraints on the time of the first caustic crossing, restrict the search
further as discussed following equation (\ref{eqn:telim}).

	We now consider the realistic case more closely.  
Since $F_{\rm cc}$ and $F_\b$ have uncertainties, so will 
$F_s$ through equation (\ref{eqn:fseval}).  Then, the uncertainties in $F_s$ 
and $Q$ will propagate to the estimate of $\sin\phi$ through equation
(\ref{eqn:sinphieval}).  Since $Q$ and  $F_{\rm cc}$ are highly anti-correlated,
the error in  $\sin\phi\propto F_s^2/Q$ will be higher than given by naive
error propagation.  One then needs to explore a range for $\sin\phi$
(say 2 or 3 $\sigma$) rather than the single value derived in the previous
section.

	Usually, the uncertainty in $F_\b$ will lie at one of two extremes.  
Either the baseline is very well known from many observations before or
after the event, or it is very poorly known because the event is not yet
over.  In the latter case, there will of course be baseline measurements
made using the telescope from which the event was discovered, but these
may not be generally available.  Even if they are, they will usually be
in a different filter with different seeing conditions and so not
directly useful for establishing a baseline for the observations of the
caustic crossing (but see \S\ 5.3).  In the first case, the error in $F_s$ is 
simply
$(A_{\rm cc} -1)^{-1}$ times the error in $F_{\rm cc}$.  In the second case, one has
only an upper limit, $F_\b < F_\lim$.  This leads to range of allowed values
for $F_s$,
\begin{equation}
{F_{\rm cc}-F_\lim\over A_{\rm cc} - 1}\leq F_s \leq {F_{\rm cc}\over A_{\rm cc}},\label{eqn:fsrange}
\end{equation}
with the second relation coming from $A_{\rm cc} F_s = F_{\rm cc} - F_b \leq F_{\rm cc}$.  This
range must then be expanded to allow for errors in $F_{\rm cc}$ before being
combined with equation (\ref{eqn:sinphieval}) and its associated uncertainties
in $Q$.

	Once the four trial parameters $(d,q,\ell,\te)$ are chosen, the
allowed range in $\phi$ can be determined.  
The standard parameters $t_0, u_0$, and $\alpha$, which completely 
specifying the trajectory can be found from the relations in Table 1,
and using $d$, $q$, and $\te$, the magnification can be determined
as a function of time.
The best fit for the remaining two parameters needed for the non-crossing data, 
$F_s$ and $F_b$, can be determined by linear regression.  That is, for each non-caustic-crossing
observation (to be defined more precisely below) at time $t_i$, 
one predicts the flux,
\begin{equation}
F_{{\rm pred},i} = A^0(t_i)F_s + F_b, \label{eqn:fpred}
\end{equation}
and then forms $\chi^2 = \sum_i (F_{{\rm pred},i} - F_{{\rm obs},i})^2/
\sigma_i^2$, where $F_{{\rm obs},i}$ is the observed flux, and $\sigma_i$
is the error of the observation at $t_i$.  This does not yet take into account
the information about $F_s$ and $F_b$ contained in the caustic-crossing
data.  To include this information, we simply invert equation 
(\ref{eqn:sinphieval}) and note that in the present context, $\phi$ and $\te$
should both be regarded as constants.  That is, 
$F_s = [Q\sin\phi/(u_r\te)]^{1/2}$
and $\sigma_{F_s} = \sigma_Q[\sin\phi/(4 Q u_r\te)]^{1/2}$, where 
$\sigma_Q$ is the
error in $Q$ taken from fit to the caustic-crossing data.  Hence, $\chi^2$
is given by,
\begin{equation}
\chi^2 = \sum_i {(A^0(t_i) F_s + F_b - F_{{\rm obs},i})^2\over\sigma_i^2} 
+  4\,{[(Q u_r \te\csc\phi)^{1/2} F_s - Q]^2\over\sigma_Q^2},\label{eqn:chisq}
\end{equation}
which can be solved for $F_s$ and $F_b$ by standard linear techniques.

	Clearly all the points that were not used in the caustic-crossing
fit can be incorporated into equation (\ref{eqn:chisq}).  In addition, one
might also wish to use the points outside the caustic which were included
in the caustic-crossing fit in order to determine $F_{\rm cc}$ and the slope 
$\tilde \omega$.  Since $F_{\rm cc}$ does not directly enter equation (\ref{eqn:chisq}),
this may appear to be permissible.  Actually, since $F_{\rm cc}$ is highly
correlated with $Q$, inclusion of these points is not strictly permitted.
Nevertheless, we advocate including them (and thus slightly overcounting
the information they contain) because the method is being used to find
allowed regions of parameter space, not to determine the errors of the best
fit.

\section{A Worked Example: PLANET data for MACHO 98-SMC-01}

	To illustrate how the method works, we apply it to the PLANET data
for MACHO 98-SMC-1.  These data differ from those analyzed by
Albrow et al.\ (1999a) in two ways: the SAAO data have
been re-reduced using a better template, and a few late times points that 
became 
available only later have been added.  In addition, we now report the Heliocentric Julian
Date (HJD) rather than Julian Date (JD) and uniformly report the midpoints
of the exposures, rather than their beginnings as was previously done for some
of the observatories.  We choose this example because it has a well-covered
caustic and the data are publically available 
(http://www.astro.rug.nl/$\sim${planet}).  

\subsection{Choosing the Data Set for the Caustic-Crossing Fit}

	The first step is to fit the caustic crossing, and to do this we must
choose which data points should be used for the fit.  The entire data
set is shown in Figure \ref{fig:one}.  Data within 1.5 days of the caustic
crossing (HJD-2450000.0$=982.6\pm 1.5$) are shown as individual points while the rest
are shown as daily averages.  Figure \ref{fig:two} shows the immediate
neighborhood of the crossing in more detail.

	What should be the first point included in the caustic-crossing fit?
When the source is too close to the caustic, it cannot be approximated as a 
point source, and so cannot be included in the non-caustic-crossing fit.
Hence, these observations
should be included in the caustic-crossing fit.  This condition
can be understood precisely because, from equation (\ref{eqn:atrue}), $G_0(\eta)$
can be expanded in the limit $\eta\ll -1$,
\begin{equation}
G_0(\eta) = (-\eta)^{-1/2}\biggl(1 + {3\over 32}\eta^{-2} + \ldots\biggr),
\qquad (\eta<-1).\label{eqn:getaapprox}
\end{equation}
Hence, for typical daily-averaged photometry errors of $\la 1\%$, this cut
off should be about 3 source-radius crossing times before the
time of the caustic crossing, i.e., at $t=t_{\rm cc}-3\Delta t$
where the fractional effect of the finite source is
$(3/32)\eta^{-2}\sim 1\%$.  For well sampled
crossings, one can estimate $\Delta t$ and $t_{\rm cc}$ by eye, and use these
estimates to determine which data should be included.  
As we show below, for this data set $\Delta t\sim 0.18$
days, and $t_{\rm cc}\sim 982.62$ days,  so data after $t\sim 982.08$ should be
included.  Another important consideration is that
the magnifications too far before the crossing 
will not be well approximated by equation (\ref{eqn:atrue}), primarily
because the two divergent images will not be well approximated by
equation (\ref{eqn:adiv}).  If the time that the source spends inside the 
caustic is not long compared to $\Delta t$, then this condition cannot
be satisfied simultaneously with the previous one, and the method breaks
down.  This might happen either because the caustic is very small (e.g., a 
planetary caustic) or because the crossing is close to a cusp.  Other methods
must then be used (e.g., Gaudi \& Gould 1997; Albrow et al.\ 1999b).   
 From Figure \ref{fig:one}, however, the ratio of these 
two times is at least 50 in the present case, so this is not a major concern.

	What should be the last point included?  Sufficient data
after the crossing are required to establish the slope $\tilde \omega$ well enough
to extrapolate back ``under'' the high magnification peak 
($t_{\rm cc}\pm \Delta t$) and so establish the value of $F_{\rm cc}$.  In the present case,
the three Yale points near 982.8 days are too close to the end of the
caustic crossing for this purpose.  The next set of points are the SAAO data
near 983.6 days.  Fortunately, the cusp-approach ``bump'' centered near
988 days is sufficiently far from these SAAO observations
that they can be used.  In general, one might not be so lucky, and the choice
of a final cut off for data to be included in the caustic-crossing fit should
be made carefully.

	Altogether, there are 74 data points in caustic crossing region,
71 from SAAO and 3 from Yale.  Since these data come from two different
telescopes, they could in principle have different values of $F_s$ and
$F_b$.  Since the 
caustic crossing itself does not possess sufficient information
to determine the relative values of $F_s$ and $F_b$, either external
information must be applied or data from one of the observatories must
be ignored.  The latter choice would be tolerable in the present case because
there are only 3 Yale points and, as we will show, these reduce the error
bars of the caustic-crossing parameters by only about 20\%.  In general, 
however, fitting the crossing may depend critically on data from several
observatories.  Even in the present case, using all the data would be
preferable.  To do so, we first make an initial educated guess as to the 
relative values, namely that $F_s$ and $F_b$ are both the same for the
two observatories.  (In the present case, $F_s$ is known a priori to be
the same because the two observatories use similar filters and the photometric
measurements are made relative to the same reference stars.  However, the
two $F_b$ could be different because the reductions are not carried out
with the same template, and so different amounts of background light could
enter the photometric apertures.)  We then search for solutions using the 
non-caustic-crossing data.  We find that all viable solutions have
$F_b^{\rm SAAO} \simeq F_b^{\rm Yale} + 0.039 F_{20}$ where $F_{20}$ is the
flux from an $I=20$ star.  The scatter $(0.04 F_{20})$ in these determinations 
is smaller
than the $0.08 F_{20}$ combined error for the three Yale measurements, so
we simply employ the offset without incorporating an additional uncertainty.

\subsection{Caustic-Crossing Parameters}

	We find fit parameters,
\begin{equation}
Q= (15.73\pm 0.35)\, F_{20}^2\,\day,\quad
t_{\rm cc} = (982.62439\pm 0.00087)\,\day,\quad
\Delta t = (0.1760\pm 0.0015)\,\day,
\label{eqn:parmfit1}
\end{equation}
\begin{equation}
F_{\rm cc} = (1.378\pm 0.096)\,F_{20}\,\qquad
\tilde \omega = (0.02\pm 0.10)\,F_{20}\,\day^{-1},
\label{eqn:parmfit2}
\end{equation}
with a matrix of coefficients of local correlation 
\begin{equation}
\left(\matrix{
   1.00 &  0.45 &  0.64  & -0.97 &  0.91 \cr     
   0.45 &  1.00 &  0.76  & -0.39 &  0.32\cr     
   0.64 &  0.76 &  1.00  & -0.57 &  0.52\cr     
  -0.97 & -0.39 & -0.57  &  1.00 & -0.93\cr     
   0.91 &  0.32 &  0.52  & -0.93 &  1.00}
\right), \label{eqn:corcoef}
\end{equation}
where the order of the rows and columns corresponds to the order of
the parameters in equations (\ref{eqn:parmfit1}) and (\ref{eqn:parmfit2}).
The effective rise time of the caustic crossing is $t_{\rm r,eff}=
(8.28 \pm 1.34) \day$.
Note that the midpoint of
the first Yale data point occurs 4 minutes before the best-fit time for
the end of the crossing.  Since the exposure time was $t_{\rm exp} = 20\,$min,
we use equation (\ref{eqn:finiteexp}) for this point.

	For completeness, we note that if we ignore the Yale data, we obtain
$Q= (15.73\pm 0.41)\,F_{20}^2\,\day$, 
$t_{\rm cc} = (982.62444\pm 0.00096)\,\day$, 
$\Delta t = (0.1761\pm 0.0017)\,\day$, 
$F_{\rm cc} = (1.379\pm 0.115)\,F_{20}$, and 
$\tilde \omega = (0.02\pm 0.12)\,F_{20}\,\day^{-1}$.  

	Figure \ref{fig:two} shows the best-fit curve to the caustic crossing.
It has $\chi^2=113$ for 69 degrees of freedom.  We therefore estimate that
the formal DoPHOT errors should
be multiplied by $(113/69)^{1/2}=1.28$, and we use these higher
errors in all subsequent work.  This ratio between formal errors and true
uncertainties is typical of DoPHOT reduced PLANET data (Albrow et al.\ 1998).

We note in passing that the end of caustic crossing occurred at
$t=t_{\rm cc}+\Delta t = 982.8004\pm 0.0028$. 
This may be compared to the values
obtained by EROS from their detailed observations of the end of the
caustic crossing, $982.7987\pm 0.0012$ and $982.7997\pm 0.0021$ for their
blue and red filters respectively (Afonso et al.\ 1998), where we have
converted the EROS numbers from JD to HJD.

\subsection{Grid of Lens Parameters}

	In principle, the lens could have any geometry $(d,q)$, with
$0<d<\infty$ and $0<q\leq 1$.  We must therefore choose a grid of geometries
that adequately samples this space.  We initially choose arrays
of values $q=0.005$, 0.01, 0.03, 0.05, 0.1, 0.3, 0.5, 0.75, 1.0 and 
$d=0.3$, 0.4, 0.5, 0.6, 0.7, 0.8, 0.9, 1.0, 1.2, 1.4, 1.7, 2.0, 2.5, 3.0, 3.5, 4.0, 4.5.  
We will see in \S\ 4 that this is adequate.  For 
the Einstein crossing times, we choose
a range $20\,\day \leq \te \leq 200\,\day$.  Our observations display 
significant structure
for at least 25 days beginning at about the minimum
of the caustic region, so it is very unlikely that the
event could be shorter than 20 days.  In fact the event could be longer
than 200 days if it were heavily blended, in which case only the inner,
highly-magnified portions of the Einstein ring would give rise to significant
structure.  In this case, we would find that for each geometry
near the geometry characterizing the actual event, 
the lowest $\chi^2$ fit would have durations at or near
our upper limit of $\te=200\,$days.

	For each geometry, we first create a very densely-sampled
representation of the caustic using the algorithm of 
Witt (1990), which is
unevenly sampled with much wider spacing near the cusps than between them.  
We then resample each of the 1 to 3 closed caustic curves with about 800
roughly equally spaced points, $\ell_i$.  
At each point we evaluate $A_{\rm cc}(\ell_i)$ directly
on the caustic and $u_{r}(\ell_i)$ by sampling the magnification at
distances $\Delta u_{\perp}=0.0001,0.00004,0.00002,$ and $0.00001$ and applying
equation (\ref{eqn:adiv}). This procedure of course fails in the
neighborhood of the cusps, but since the largest trial value of $\Delta u_{\perp}$
is more than 10 times smaller than the source, any caustic position where the
procedure fails is not a viable candidate for a fold caustic crossing in any case.

	The event was not yet at baseline
at the last data point.  We therefore can
estimate only an upper limit $F_\lim>F_\b$ for the baseline flux.  We choose
$F_\lim = 0.55\,F_{\rm 20}$ based on the upper limit from the last three
measurements (see Fig.\ \ref{fig:one}).

	We find that stepping through the $\sim 800$ caustic points $\ell_i$ 
yields a $\chi^2_{{\rm min}}(\ell_i)$ as a
function of position $\ell_i$ that is sufficiently well sampled to obtain
at least one point with $\chi^2_{\rm min}$ that is within 1 or 2 of the
true local minimum.

	At each position we sample the range of time scales $\te$ in
increments of 5\%.  This choice is dictated by the character of the
cusp-approach structure seen in Figure \ref{fig:one} with a peak near
$t=988$ days.  The full width at half maximum and the time from the caustic
crossing are about equal, approximately 6 days.  The 14 daily-averaged 
measurement errors are typically $\sigma\sim 4\%$.  Hence a deviation of the 
trial $\te$ from the true value by $\delta\sim 2.5\%$ 
would lead to a change in  $\chi^2$ of $\sim 14(\sigma/2\delta)^2\sim 1$.
For each $\te$, we explore the range of $\sin\phi$ described by 
equations (\ref{eqn:sinphieval}) and (\ref{eqn:fsrange}), and augmented
by the $3\,\sigma$ errors for $Q$ (eq.\ \ref{eqn:parmfit1}), stepping
in 5\% increments.  Other choices of increment size could be made.
Empirically we find that with our adopted choice of 5\% timescale increments, 
the search can miss a local minimum in $\chi^2$ by
$\Delta \chi^2\sim 10$.  This means that all local minima lying within 
$\Delta \chi^2\sim 15$ of the global minimum must be checked (see \S\ 4).

	For each geometry $(d,q)$ we record the lowest value of $\chi^2$ and
examine the resulting map.  We find three very broad areas of $(d,q)$ space
with very similar values of $\chi^2\sim 130$--135.  These are roughly described
by $(0.4 \la d \la 0.7)\times(0.3 \la q \la 1)$, 
$(2.5 \la d \la 3.5)\times(0.1 \la q \la 1)$, 
and
$(0.6\la d \la 0.7)\times(0.05 \la q\la 0.1)$. That is, we appear to have found
an extremely broad class of solutions rather than a single unique minimum
or even a few well-defined isolated minima.  

	For several individual $(d,q)$ pairs, we also examine the minimum
$\chi^2$ {\it as a function of position} $\ell$ around the caustics.  
Typically, we find two distinct minima with comparable $\chi^2$,
one with $\alpha$ close to 0 (or $2\pi$) and the other with 
$\alpha\sim \pi$.  These describe second caustic crossings on opposite
sides of the caustic region.  We therefore conduct two automated searches at
each $(d,q)$, one with $\pi/2\leq\alpha<3\pi/2$ and the other in the
complementary region.  We remark on the relation between these two solutions
at the end of \S\ 4.

\section{Worked Example II: Refined Search for Minima}

	In order to investigate this preliminary result further, we search 
for local minima near the solutions with $\chi^2 \la 145$, 
found at each $(d,q)$ 
encompassing a slightly larger region than the 
broad apparent plateau discussed at the end of the previous section.  We adopt
this somewhat looser criterion because, as we discussed above, the initial
systematic search could miss the true minimum by $\Delta\chi^2\sim 10$.

\subsection{Basic Approach}

	Although the standard procedure in such a search would be to allow
all parameters to vary simultaneously, we specifically {\it do not} follow
this usual approach.  Instead, we hold $d$ and $q$ fixed and allow only 
the remaining parameters to vary.  
This will permit a test of the hypothesis that
there are a set of very broad minima in $(d,q)$ space.  If the $\chi^2$
minimum at each of these points is essentially the same, then $d$ and $q$
are indeed highly degenerate.  On the other hand, if the minimum
$\chi^2$ is found to differ substantially for different fixed $(d,q)$, then
it would be worthwhile to allow these parameters to vary simultaneously with
the others.

	For each set of trial parameters, we proceed as follows.  For each 
observation (not binned by day as in the preliminary search), we evaluate the
magnification by one of two methods, both semi-analytic.  If the source
lies entirely outside of the caustics or if its center lies at least
3.5 source radii from a caustic, we simply use the magnification at the source
center.  Otherwise, we use an approximation for the magnification that is
similar in spirit to the approximation used to fit the caustic crossing that we
introduced in \S\ 2.1,
\begin{equation}
A(\bu_p) = 
yA_3^0 ({\bf u}_p) + A_2^0({\bf u}_q)\biggl({\Delta {u}_{q,\perp}\over\rho_*}
\biggr)^{ 1/2}
G_0\biggl(-{\Delta {u}_{p,\perp}\over \rho_*}\biggr),
\label{eqn:analytica}
\end{equation}
where ${\bf u}_p$ is the position in the Einstein ring of the center of the
source, ${\bf u}_q$ is another position in the Einstein ring to be described
below, $\Delta {u}_{p,\perp}$ and  $\Delta {u}_{q,\perp}$ are respectively the 
perpendicular distances from ${\bf u}_p$ and
${\bf u}_q$ to the nearest caustic, 
$A_3^0 ({\bf u}_p)$ is the magnification of the 3 non-divergent images
at the position ${\bf u}_p$, $A_2^0({\bf u}_q)$ is the magnification of the
2 divergent images at the position ${\bf u}_q$,
and $\rho_*$ is the source size in units of the Einstein ring.  If
$\Delta {u}_{p,\perp}>\rho_*$, then we assign ${\bf u}_q= {\bf u}_p$.
Otherwise, we take ${\bf u}_q$ to lie along the perpendicular to the caustic
through ${\bf u}_p$ and halfway from the caustic to the limb of the star
that is inside the caustic.  The argument of $G_0$ is negative if the center 
of the source lies inside the caustic and positive if it lies outside.  

	Note that for the second term in equation (\ref{eqn:analytica}) to be 
well defined, the arguement of $A_2^0$ {\it must} be a point inside the 
caustic.  This is the reason for choosing a $\bu_q$ different from $\bu_p$.  
If the  approximation given by equation (\ref{eqn:adiv}) (and so eq.\ 
\ref{eqn:abu}) were exact, then
equation (\ref{eqn:analytica}) would be valid with {\it any} choice of
$\bu_q$ inside the caustic.  
Since equation (\ref{eqn:adiv}) is not exact, we choose $\bu_q$
at the middle of the part of the source inside the caustic in order to
minimize the error.

	As we discussed in \S\ 2, this approximation should work well whenever
the source is small compared to the distance between caustic crossings and to
the distance from a caustic crossing to the nearest cusp.  It will not work
for small (e.g.\ planetary) caustics or cusp crossings.

	The advantage of this approximation is that it allows one to evaluate
the magnifications for the several hundred points on the light curve in less
than one second, compared to several minutes required for a numerical
integration over the source.  This advantage will come into play when we
discuss our minimization technique below.

	Once the magnifications have been calculated we fit for the flux
parameters.  Recall that for a single observatory there are two parameters,
$F_s$ and $F_b$.  In this example, there are four observatories,
Canopus 1m, CTIO 0.9m, SAAO 1m, and Yale 1m, which seems to imply
8 flux parameters.  However, since all four observatories use very similar
$I$ band filters and reduce the images relative to a common set of 
local standards, we take $F_s$ to be the same for all four.  In addition,
we take the $F_b$ for Canopus to be the same as SAAO because there is only
one data point (and so no room for another parameter) and because it is
a high magnification point so differences in $F_b$ are unlikely to be 
important.  The linear fit to the remaining 4 parameters 
requires very little time to compute (see also Rhie et al.\ 1999).

	Since $d$ and $q$ are held fixed, and $F_b$ and $F_s$ are determined
by linear regression, there remain 5 parameters to
fit.  These are normally taken to be $t_0$, $u_0$, $\te$, 
$\alpha$, and $\rho_*$.  However, as discussed in \S\ 2, the time of
the caustic crossing, $t_{\rm cc}$, and the caustic-crossing time, $\Delta t$,
are much better determined than either $t_0$ or $\rho_*$; we therefore
use the former in place of the latter.  While both $t_{\rm cc}$ and $\Delta t$
are allowed to vary, both tend to move over very small ranges that
are consistent with the results from the caustic-crossing fit in
\S\ 3.2.  Nevertheless, despite the fact that two parameters are held fixed
and two others are relatively well constrained, we find that it is not easy
to locate local minima.  We suspect that the $\chi^2$ function is quite
complicated.  Moreover, in order to properly explore parameter space, it
is necessary to repeat the minimization procedure for several dozen different
$(d,q)$ pairs, and this will be multiplied several fold in the next section.

	We therefore take advantage of the efficient method of magnification
calculation summarized in equation (\ref{eqn:analytica}), which makes possible 
a rather cumbersome, but fairly robust, method of minimization.  For each
parameter $a_i$ we establish a grid size, $\delta_i$, and for every new
set of trial parameters evaluate $\chi^2$ at the 51 positions
$(a_i + \epsilon_i\delta_i)$, where $\epsilon_i= -1,0,+1$ and
$\sum_i|\epsilon_i|\leq 2$.  At each step, the operator is allowed one
of three options: move to the lowest value of the 51 positions, move
to (or toward by a specified fraction) the predicted minimum of the best
fit quadratic to the $\chi^2$ surface, or adjust the grid size.  In practice,
the procedure is semi-automated so as not to bother the operator while
the routine is making adequate progress.

	We find that
even with this extensive probing of the neighborhood of the trial solution,
the path to lower $\chi^2$ is not always apparent.  For example, sometimes
none of the 50 probes of parameter space has a lower $\chi^2$
than the central position, even if the grid size is
decreased by a factor 2 or 4.  We then move toward the best estimate of the minimum
that is derived from the quadratic fit and find that this also has higher
$\chi^2$.  However, starting from this new central position, some of the
50 new probes have substantially lower $\chi^2$.  Moreover for the next
iterations the path downward is clear, and $\chi^2$ may drop by 2--10 over
these next few steps.  We do not understand the nature of these
``hang-ups.''  In principle, it is possible they are due to genuine local
minima, but we suspect that the $\chi^2$ surface
is just extremely complicated and that the paths toward lower values are 
narrow and not well probed even by our 50 trial points.  Although
skepticism is warranted, we
believe that the true local minimum is eventually reached,
for two reasons.  First,
as we show in the next section, we find many different solutions with almost
exactly the same $\chi^2$.  It would be a remarkable coincidence if
the search process always stalled at the same value of $\chi^2$.
Second, if the first attempt does not
approach this minimum, we try several other ``paths'' and we find that
there are no significant improvements after the second or third try.
Nevertheless, this experience counsels us to be cautious about the
interpretation of apparent minima.

\subsection{Solutions}

	We search for refined solutions (see \S\ 4.1) near each of the
rough solutions found in \S\ 3 considering only those within 
$2.5\,\sigma$ (i.e., $\Delta\chi^2<6.25$) 
of the minimum value found for the entire grid.  
We find 41 such solutions including {\it all} combinations
of $d = (0.4,0.5,0.6,2.5,3.0,3.5)$ and $q = (0.3,0.5,0.75,1.0)$, plus
additional solutions at $(d,q)=(0.6,0.1)$, (0.7,0.05), (0.7,0.1), and
(0.7,0.3).  This
appears to be only 28 solutions, but for many $(d,q)$ pairs we find two
distinct solutions, one at $\alpha\sim 0$ (or $2\pi$) and the other at
$\alpha\sim \pi$.  

Table 2 shows the 41 solutions.  The first
seven columns are the parameters $d,q,\alpha,u_0,\te,t_{\rm cc}$, and $\Delta t$.  
The next two are the $x$ and $y$ components of ${\bf u}_{\rm cc}$, the point of the
caustic crossing.  These are shown in order to allow easy transformation
into other parameterizations of the geometry.  Columns 10 and 11 show
$F_s$ and $F_b$ (from SAAO), and column 12 is their sum, $F_\b=F_s+F_b$,
expressed as a magnitude, $I_\b = 20 - 2.5\log F_\b$.
Column 13 is $t_*\equiv \Delta t \sin\phi$.  Recall that the proper motion
is given by $\mu = \theta_*/t_*$.  Finally, column 14 is $\Delta\chi^2$, 
defined by,
\begin{equation}
\Delta\chi^2 \equiv {\chi^2 - \chi^2_\min\over \chi^2_\min/\rm dof},
\label{eqn:deltachitwodef}
\end{equation}
where $\chi^2_\min = 467.95$ is the minimum value of $\chi^2$ found in our
search at $(d,q)=(0.5,0.3)$, and $\rm dof = 212 - 11 = 201$ is the number
data points minus the number of parameters.  
(Note that the fluxes are calibrated to the Cousins system based on
comparison to OGLE stars.  See Albrow et al.\ 1999a.)

	The basic result illustrated by Table 2 is that a broad range of
parameters are permitted by the data.  Two very broad regions in 
($d,q$) space, one with $d<1$ and the other
with $d>1$ are permitted. Indeed, there is a rough symmetry 
$d\leftrightarrow d^{-1}$ which was theoretically predicted by Dominik (1999b).
There is also a third, smaller region centered at $(d,q)\sim (0.7,0.1)$.
As expected, $t_{\rm cc}$ and $\Delta t$ lie in an extremely narrow range since they
are primarily determined by the caustic structure and not the global 
parameters.  The full solutions for both $t_{\rm cc}$ and $\Delta t$ deviate by
about $0.0015$ days ($\sim\, 2$\, minutes) from the caustic-crossing solution given in equation
(\ref{eqn:parmfit1}).  This shows that the global
parameters do have some influence on the determination of the caustic-crossing
parameters, although it is quite small.  More striking is the large variation
in permitted Einstein crossing timescales $\te$, from 81 to 227 days.
Also of note is the wide variation in allowed values of $t_*$, from
0.70 to 3.42 hours.

	Of course, the fact that there are 41 solutions 
rather than some other number is a result of our specific choice of grid.  
Since Table 2 shows very little structure in $\Delta\chi^2$ over broad
ranges of $(d,q)$ space, we expect that a finer grid would not yield any
additional information.

	How different are the light curves associated with these various
fits?  Figures \ref{fig:three}--\ref{fig:five} show four representative
examples, two from the $d>1$ region and two with $d<1$.  Each figure 
contains a curve for  $(d,q)=(0.5,0.3)$, the nominal ``best fit,'' and
also one other, for $(d,q)=(3.5,0.75)$, $(0.7,0.05)$, and
$(0.6,0.75)$, respectively.  All four solutions have $\alpha\sim \pi$
(see Table 2).
In Figure \ref{fig:three}, the caustic-crossing region
is shown separately.  This is not done for the other two figures because the
caustic-crossing regions look identical for all four solutions.
All light curves are normalized to the SAAO data
by subtracting
$\Delta F = F_{b,i} - F_{b,\rm SAAO}$, i.e. the difference in the fit
values for the backgrounds as measured at the two observatories.  In each
case, the two light curves are barely distinguishable over the time period
covered by the data.  This shows that a wide variety of geometries can
produce essentially identical light curves if one is restricted to data
covering the ``second half'' of the event.  On the other hand, in the regions
that are not covered, the light curves can differ dramatically.

	An important corollary to this observation is that, by time reversal,
it is impossible to accurately predict the time of the second caustic
crossing even from extremely good data covering the first.  The second
caustic crossing can only be predicted by frequent monitoring of the event
and looking for the inverse square root behavior as the second caustic
approaches.  Indeed, this is how the second caustic crossing of MACHO 98-SMC-1
was predicted by PLANET; the predictions of MACHO close to the caustic crossing
were made in this way as well.

	It is interesting to examine the relation between the two solutions
with the same $(d,q)$.  From Table 2, one finds that these generally
have similar times scales $t_{\rm E}$ and angles $\alpha$ that differ
by approximately $180^\circ$.  However, the caustic crossing angles
$\phi$ can be quite different.  (Since $\sin\phi=t_*/\Delta t$, and
$\Delta t$ is essentially identical for all solutions, 
$\sin\phi\propto t_*.$)\ \ 
For $d<1$, these differences are severe for small value of $q$ and
diminish at $q\rightarrow 1$.  This behavior can be understood by
examining Figure \ref{fig:onep}: since $\alpha$ changes by about $180^\circ$
and $u_0$ remains similar for the two soutions, the trajectory followed in
the second solution is roughly the reverse of the first.  For $q\rightarrow 1$,
the caustic becomes symmetric, so the angles of the first and second
caustic crossings become the same.  For $q$ different from 1 (e.g.
$q=0.3$ as in Fig.\ \ref{fig:onep}) the caustic is asymmetric, so the two
angles are different.  This reasoning does not apply to the $d=3.5$ solutions
because the trajectories are not approximately time reversals of each other.

\section{Additional Observations to Break Model Degeneracies}

	Since the PLANET data set covers only a portion of the light curve
(albeit very well), one might well ask what additional observations would
be required to break the degeneracies presented in Table 2.  The light
curves of the fits at each $(d,q)$ differ substantially in the regions
not covered by the data, so it might appear that even data of modest
quality in these regions would be adequate to distinguish among the various
models.  However, it is possible that for
a given $(d,q)$ there are other models in which the first caustic is at a 
different time, or the baseline flux has a different value, and while not
the absolute ``best'' fit to the PLANET data, are still compatible with it.
If this is the case, then additional data may leave
the degeneracies essentially intact.

	We therefore explore three examples of additional data that typically 
might be available: a precise measurement of the baseline, moderately
good coverage of the first caustic, and lower quality coverage of the
full light curve (including the early part) but that misses the first caustic.

	To investigate the role of additional data, we will assume that our 
``best fit''
$(d,q,\alpha)=(0.5,0.3,177^\circ)$ is in fact the true geometry.  
We emphasize that our data
cannot in fact distinguish between the various solutions shown in Table 2.
We make this assumption solely for the purpose of exploring the value of
additional data.

\subsection{Baseline}

	A year (or certainly two) after the caustic crossing, the event
will be over and a precise measurement of the
baseline can be made.  For definiteness, we will assume that this measurement
is accurate to 1\% and is taken when the event has ended.  Inspection of
Table 2 shows that $I_\b$ varies by more than 0.2 mag for the various
allowed solutions.  

If we add an additional baseline measurement and repeat the entire search
procedure, many solutions are eliminated
but 27 remain, including examples from all three regions.  In
particular, all combinations of $d = (0.4,3.5)$ and $q=(0.3,0.5,0.75,1.0)$
are allowed, as well as $(d,q)=(0.6,0.1)$, $(0.7,0.1)$, and $(0.7,0.05)$, 
and various
combinations of $d = (0.5,0.6,2.5,3.0)$ with $q=(0.3,0.5,0.75,1.0)$.
Among these solutions, $t_{\rm E}$ varies in the range 100 to
227 days, and $t_*$ varies in the range 0.90 to 3.42 hours.  A broad range
of solutions survive partly because many of the original solutions had
baselines close to that of the ``best fit'' and so were not affected by
the addition of a baseline ``measurement.''  However, a number of $(d,q)$ pairs
whose solutions shown in Table 2 would be ruled out at the $7\,\sigma$ level
by a baseline measurement, have alternative solutions that nevertheless
manage to meet the baseline constraint.  This is not true
of all solutions.  For example, the solution $(d,q)=(0.6,0.75)$ which
is shown in Figure \ref{fig:five}, did not survive the addition of a baseline
measurement.

	The broad
degeneracy in the space of solutions, even with the addition of a precise
baseline measurement, confirms the conclusion drawn at the end of \S\ 4,
that it is impossible to predict the time of the second caustic crossing
from detailed observations of the ``first half'' of the light curve.

\subsection{First Caustic Crossing}

	If the event were alerted before the first caustic crossing,
this caustic might be reasonably well covered as a result of routine
monitoring by follow-up teams.  In this case they might
notice the crossing and begin monitoring more intensively.  Nevertheless,
it is instructive to ask how well simple follow-up monitoring (i.e., without
the extra observations triggered by an anomaly alert) over the
first caustic crossing would do at resolving the degeneracies seen in Table 2.
To be specific, we assume that a total of 5 measurements are made at
equal intervals between $t_{\rm cc,1}-0.2\,$days to $t_{\rm cc,1}+0.2\,$days, and
that these have precision similar to the SAAO data at similar magnitudes,
i.e., errors of 7\%, 7\%, 1.0\%, 1.5\%, and 2\%.  Scaling from the
error estimates in equation (\ref{eqn:parmfit1}) derived from 74 data points,
these data should be sufficient to fix the time of the first caustic crossing
to $\sim 1$ hour.  By contrast, the curves shown in Figures 
\ref{fig:three}--\ref{fig:five} differ in their times of first
caustic crossing by several days.  In addition these few measurements also
strongly constrain the first-caustic crossing time (analogous to $\Delta t$)
and the scale of the first caustic (analogous to $Q$).  We find that these
few data points are sufficient to exclude all solutions found in \S\ 4,
except the assumed ``true'' solution $(d,q) = (0.5,0.3)$.

	We argued in \S\ 4.1 that the grid sampling was sufficiently fine
because $\chi^2$ was approximately flat over large contiguous regions of
the grid.  In the present case, one point on the grid has significantly
lower $\chi^2$ than all others, so this argument fails.  However, at least
for the region $d<1$, the sampling is still adequate
to find an approximate local minimum which could then act as a starting
point to find the the actual local minimum (as described in the first
paragraph of \S\ 4.1).  On the other hand, because of the generic nature
of the $d\leftrightarrow d^{-1}$ degeneracy (Dominik 1999b), one should be
cautious about claiming that there are no $d>1$ solutions simply because
there are none on the grid.  To truly rule this out, it would be necessary
to search on a much finer grid where the grid spacing was set by the
range of $(d,q)$ values around the minimum at $(d,q)\sim (0.5,0.3)$ for
which $\Delta\chi^2\la 1$.  Since we have not conducted such a search, we
cannot absolutely rule out the possibility that a $d>1$ solution survives
the addition of data from the first caustic.

\subsection{Constant Coverage}

	Next we assume that the event was covered by routine survey monitoring,
once every other day (to allow for weather) for 1000 days before the second
crossing and continuing until the end of the PLANET observations on day 1026.  
In order to complement the investigation in \S\ 5.2, we assume 
that no observations were taken within two days of either caustic crossing.
However, to take account of the fact that survey data are usually taken in
non-standard bands, we add two extra parameters to the fit, $F_s$ and $F_b$
for the survey observations.
We assume 20\% errors at baseline and that the errors scale inversely as the
square root of the flux.  

	Formally we find that only two solutions survive in addition to the
``true solution'' at $(d,q)=(0.5,0.3)$, both in its immediate neighborhood
at (0.5,0.5) and (0.6,0.3).  However we also find a cluster of
spurious solutions
centered at $(d,q)=(3.5,1)$ which has $\Delta\chi^2= 8.5$.  While it might
be possible to formally rule out such a solution in this particular case,
this low value of $\Delta\chi^2$ suggests that additional data of this
type may often leave some degeneracies intact.

\subsection{Summary}

	In brief, excellent coverage of a single fold caustic is not sufficient
to uniquely determine the parameters of the binary lens, even with the
addition of a good late-time baseline measurement.    On the other hand,
a few measurements over the other caustic can break the degeneracy
completely. This degeneracy implies that 
observations of the first caustic crossing alone cannot be used to reliably predict
the time of the second crossing. The addition of survey-type data (infrequent sampling with large
errors but covering the whole light curve -- even if the caustics are missed)
can certainly lift some of the degeneracies, but may leave the 
$d\leftrightarrow d^{-1}$ degeneracy intact.

\begin{acknowledgements}

This work was supported by grants AST 97-27520 and AST 95-30619 from the NSF, 
by grant NAG5-7589 from NASA, by a grant from the Dutch ASTRON foundation
through ASTRON 781.76.018, and
by a Marie Curie Fellowship from the European Union.  

\end{acknowledgements}

\newpage

\begin{figure}
\caption[junk]{\label{fig:flow}
Flow chart illustrating the relations among the various steps of the method
described in the text.
}
\end{figure}
\begin{figure}
\caption[junk]{\label{fig:onep}
Binary-lens parameterization.  The trajectory of the source
({\it dashed line}) is inclined by an angle $\alpha$ ($351^\circ$ in this
case) relative
to the binary axis ($M_2$ to $M_1$) ({\it solid line}).  The impact
parameter relative to the geometric center of the binary is $u_0$.
The source crosses the caustic ({\it diamond-shaped curve}) at
${\bf u}_{\rm cc}$.  The difference between this crossing and the point
of closest approach at $u_0$ is shown by a {\it bold line}.  The tangent
to the caustic makes an angle $\gamma=311^\circ$ with the binary axis.
The angle between the source trajectory and the tangent to the caustic is
therefore $\phi=\alpha-\gamma=40^\circ$. 
The vector position ${\bf u}_{\rm cc}$ is measured relative to the 
midpoint of the two masses (0,0).
}
\end{figure}
\begin{figure}
\caption[junk]{\label{fig:twop}
Binary-lens parameterization (detail).  The circle is the source which
moves on a trajectory shown by the dashed line.  The source crosses the
caustic ({\it solid line}) at position ${\bf u}_{\rm cc}$, and its current
position is ${\bf u}_p$.  The perpendicular distance to the caustic
$\Delta u_{p,\perp}$ ({\it bold line}) is used in eq.\ (\ref{eqn:analytica})
and is negative in this case.  The position ${\bf u}_q$ lies along the
line perpendicular to the caustic and half way from the caustic to the
edge of the source inside the caustic.  The distance 
$\Delta u_{q,\perp}$ ({\it bold dashed line}) is also used in eq.\
 (\ref{eqn:analytica}) and is always positive.  If the source lies 
entirely inside the caustic, then ${\bf u}_q={\bf u}_p$.  The {\it bold
crosses} show the discrete spatial sampling of the caustic 
($\sim 800$ points) in our calculations.  The vector positions
${\bf u}_{\rm cc}$, ${\bf u}_p$, and ${\bf u}_q$ are all measured relative
to the midpoint of the two masses (0,0) which is not shown in this figure
but is shown in Fig.\ (\ref{fig:onep}).
}
\end{figure}
\begin{figure}
\caption[junk]{\label{fig:one}
PLANET $I$ band data for the binary microlensing event MACHO 98-SMC-1
from four observatories, SAAO 1m ({\it crosses}), Yale-CTIO 1m 
({\it triangles}), CTIO 0.9m ({\it squares}), and Canopus 1m (Tasmania) 
({\it star}).  Individual points are shown for the interval $t=982.6\pm 1.5$.
The remaining points are daily averages at the individual observatories.
}

\end{figure}

\begin{figure}
\caption[junk]{\label{fig:two}
PLANET $I$ band data for the caustic-crossing region of the
binary microlensing event MACHO 98-SMC-1.  Fluxes are shown in units of
$F_{20}$, the flux from an $I=20$ star.  All points are from the SAAO 1m
except the three near $t=982.8$ which are from the Yale-CTIO 1m.  These
have been increased by 0.039 $F_{20}$ as described in the text.  The solid
vertical line at $t_{\rm cc} = 982.624$ is the fit value for the time that
the center of the star crossed the caustic.  The two dashed vertical lines
at $t = t_{\rm cc}\pm \Delta t$, where $\Delta t = 0.176$, are the fit values
for the time when the limbs crossed the caustic.
}
\end{figure}

\begin{figure}
\caption[junk]{\label{fig:three}
PLANET data for MACHO-98-SMC-1 together with two fits both taken from
Table 2.  In the lower panel, which shows the full light curves,
the bold curve is for  $(d,q)=(0.5,0.3)$ and the solid curve is
for $(d,q)=(3.5,0.75)$.  In the upper panel, which shows only the caustic
crossing region, the two curves are indistinguishable.  The SAAO data are
shown at the instrumental magnitudes (as reported on the PLANET web site.)
The data from the remaining three observatories are adjusted in flux by the
difference in the best fit to background flux between that observatory and
SAAO.  See text for details.  The  $(d,q)=(0.5,0.3)$ solution gives the
nominal ``best fit'' but the $(d,q)=(3.5,0.75)$ solution is worse by only
$\Delta\chi^2=0.7$.  That is, the two curves hardly differ over the region
where there are data, even though they differ drastically at earlier times.
}
\end{figure}

\begin{figure}
\caption[junk]{\label{fig:four}
PLANET data for MACHO-98-SMC-1 with fits taken from Table 2 for
$(d,q)=(0.5,0.3)$ and $(d,q)=(0.7,0.05)$.  Similar to the lower
panel of Fig.\ (\ref{fig:three}), except that $\Delta\chi^2=1.4$.  A
close-up view of the caustic crossing is not shown, since it looks
identical to the upper panel of Fig.\ \ref{fig:three}.
}
\end{figure}

\begin{figure}
\caption[junk]{\label{fig:five}
PLANET data for MACHO-98-SMC-1 with fits taken from Table 2 for
$(d,q)=(0.5,0.3)$ and $(d,q)=(0.6,0.75)$.  Similar to the lower panel
of Fig.\ (\ref{fig:three}),
except that $\Delta\chi^2=3.7$.  Note that in this case, in contrast to
the previous two, the curves deviate significantly at late times.  No
close-up view of the caustic crossing is shown because it is
identical to the upper panel of Fig.\ \ref{fig:three}.
}
\end{figure}

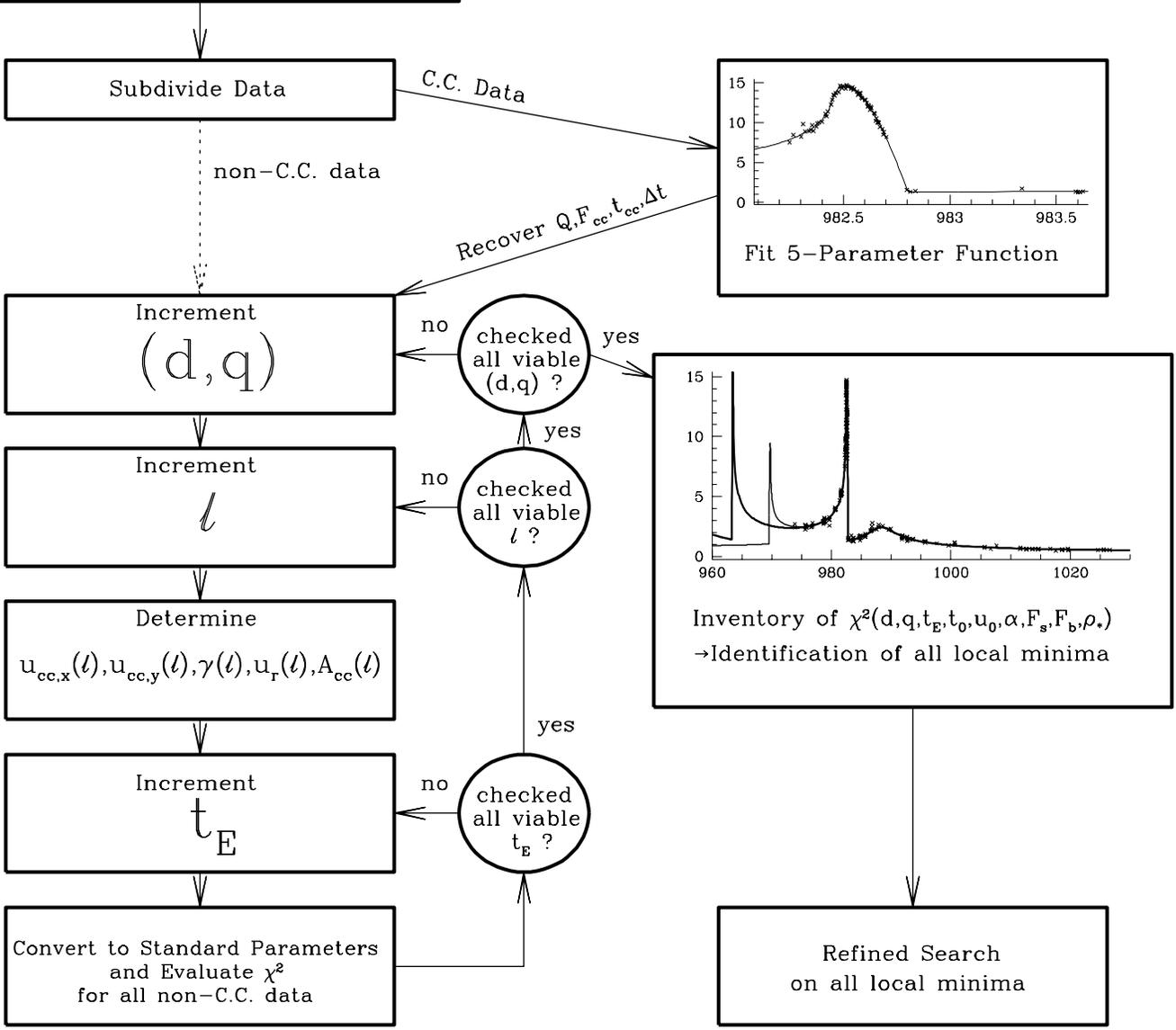
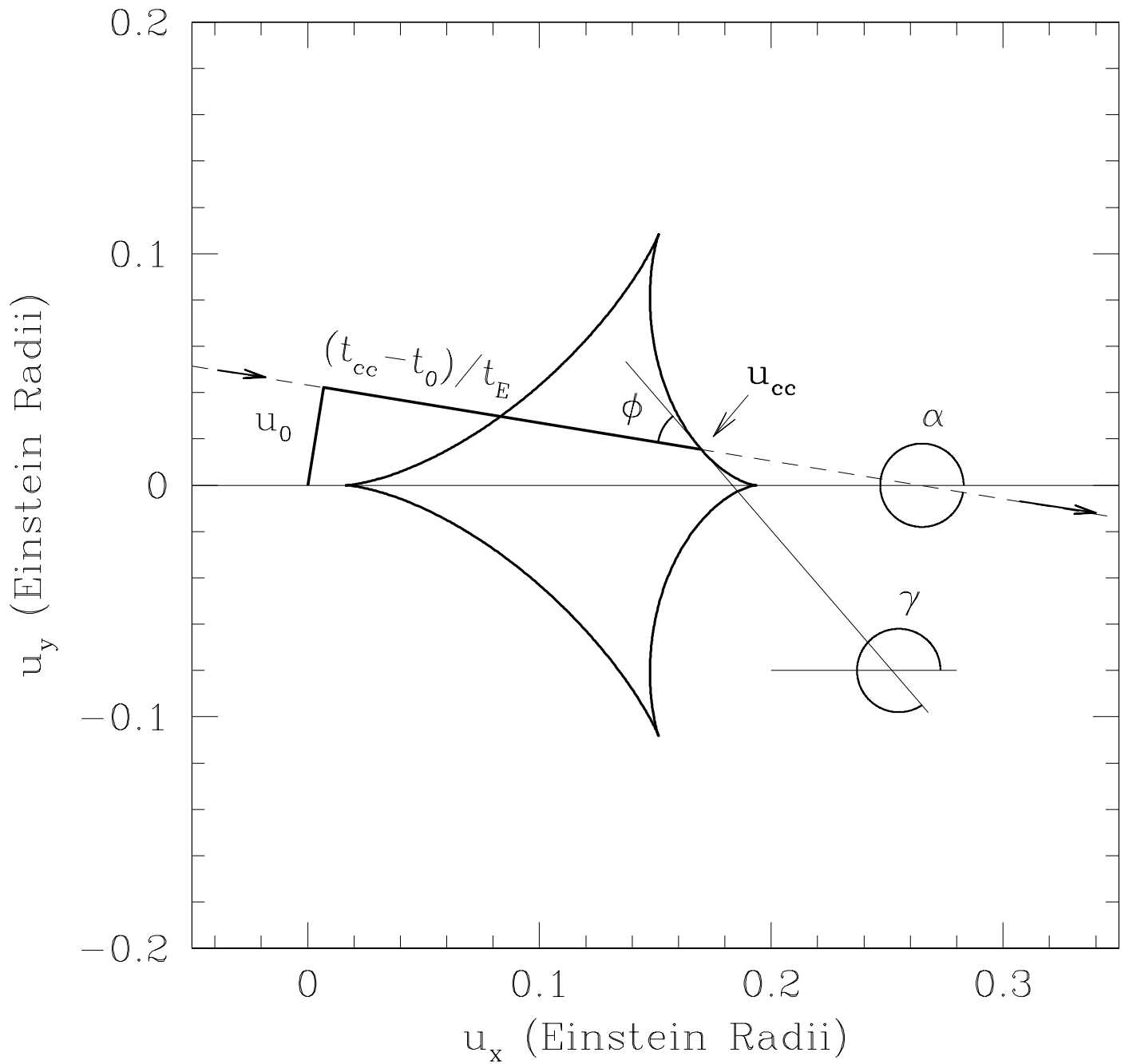
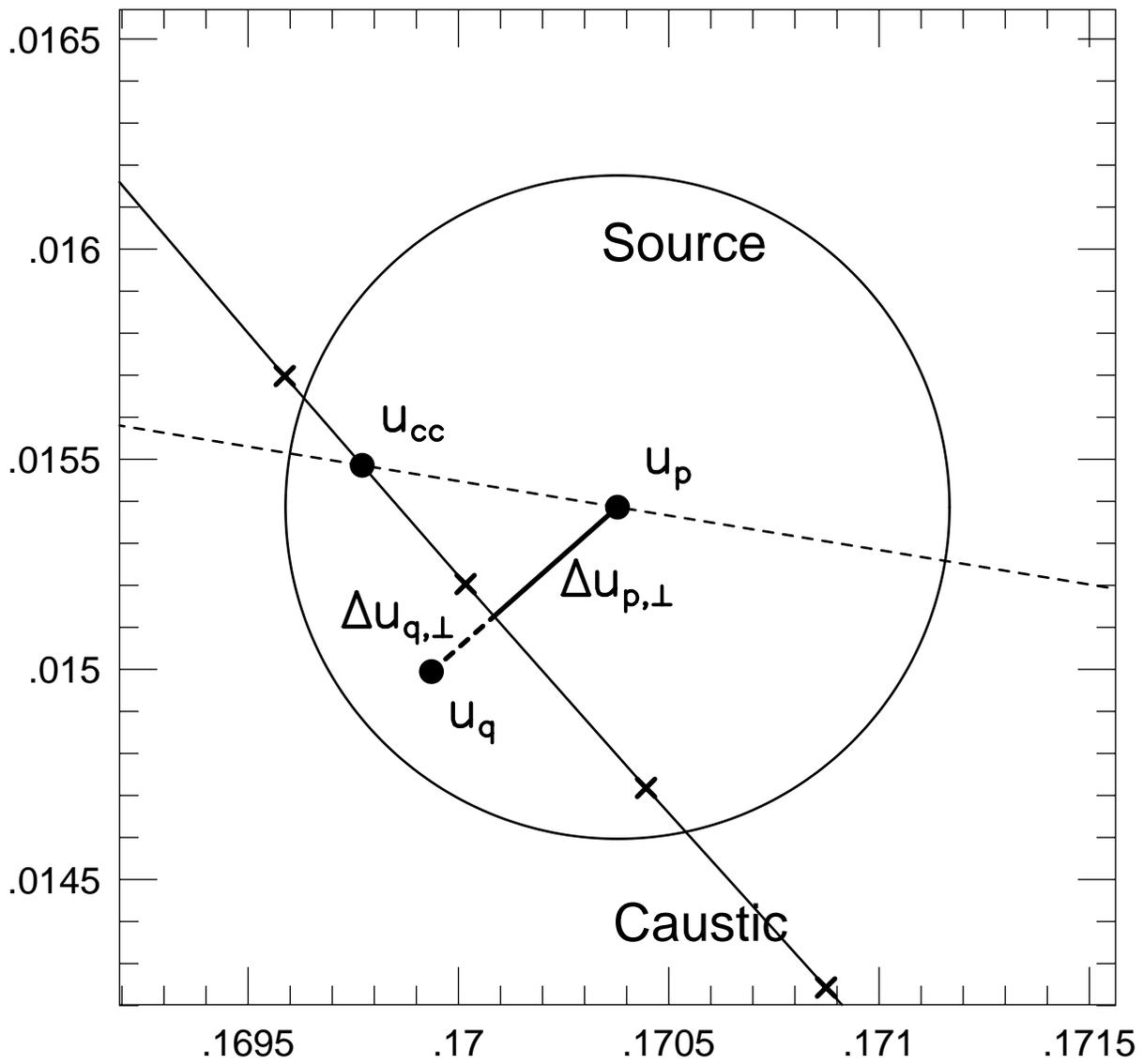
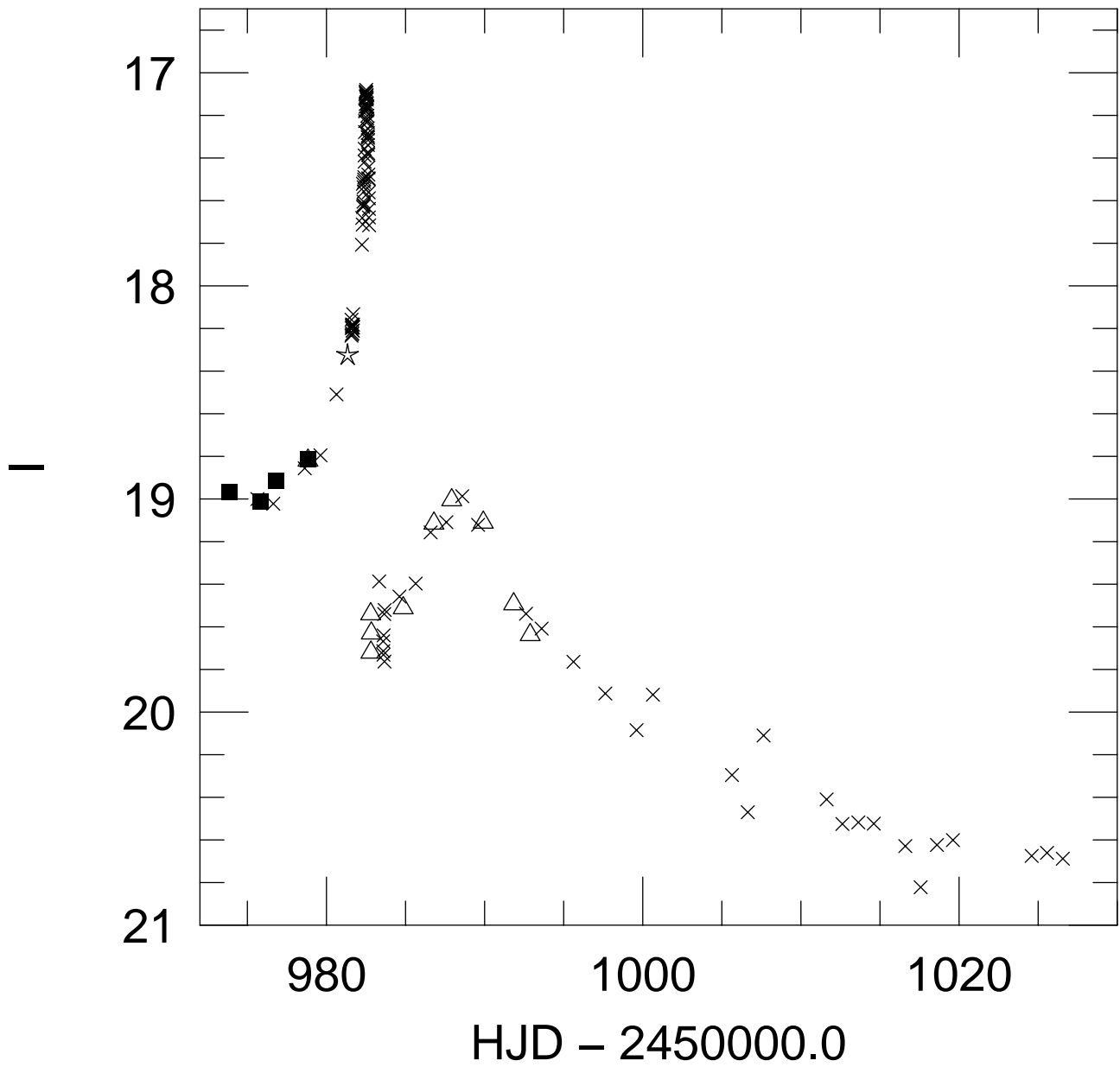
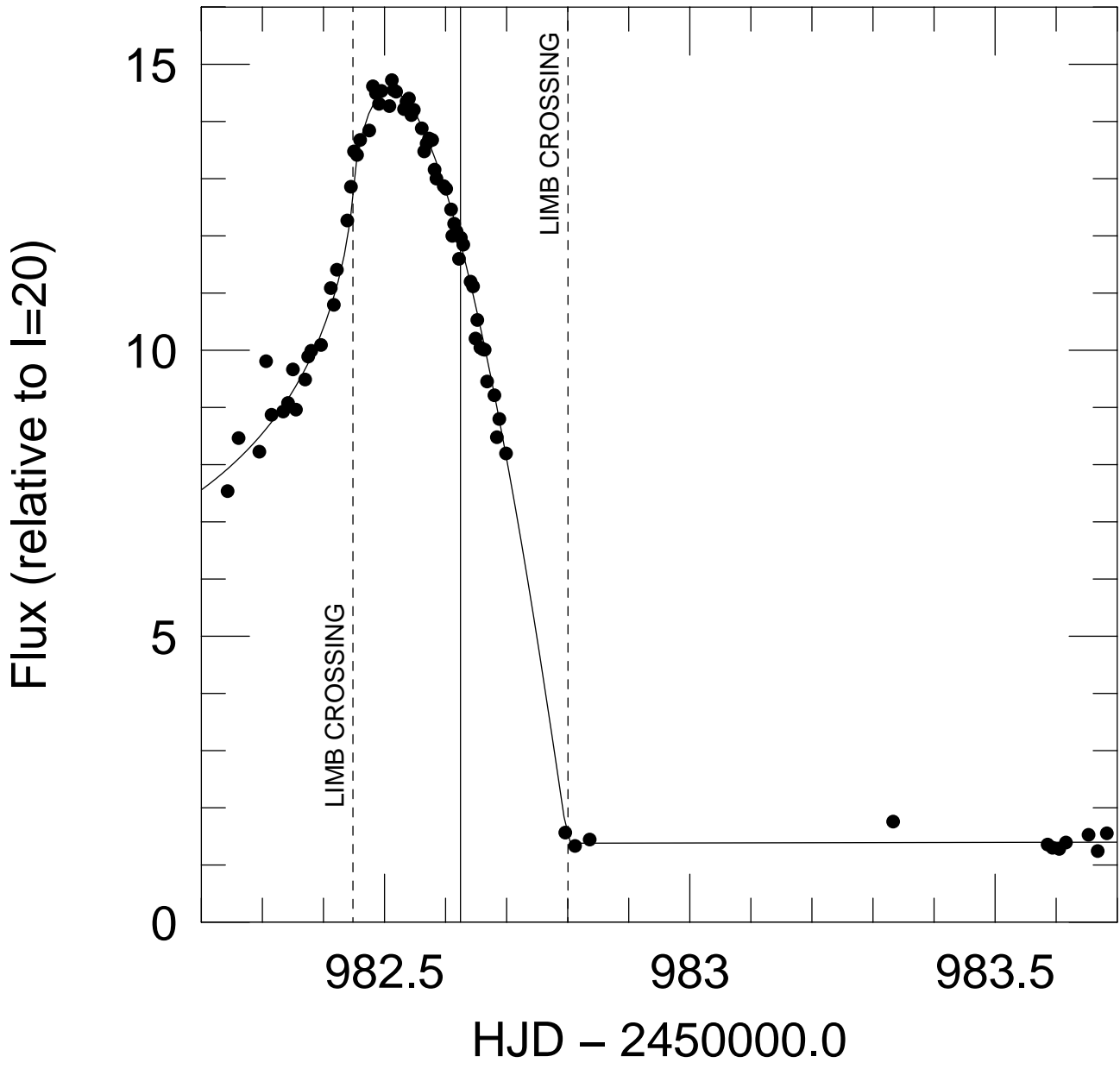
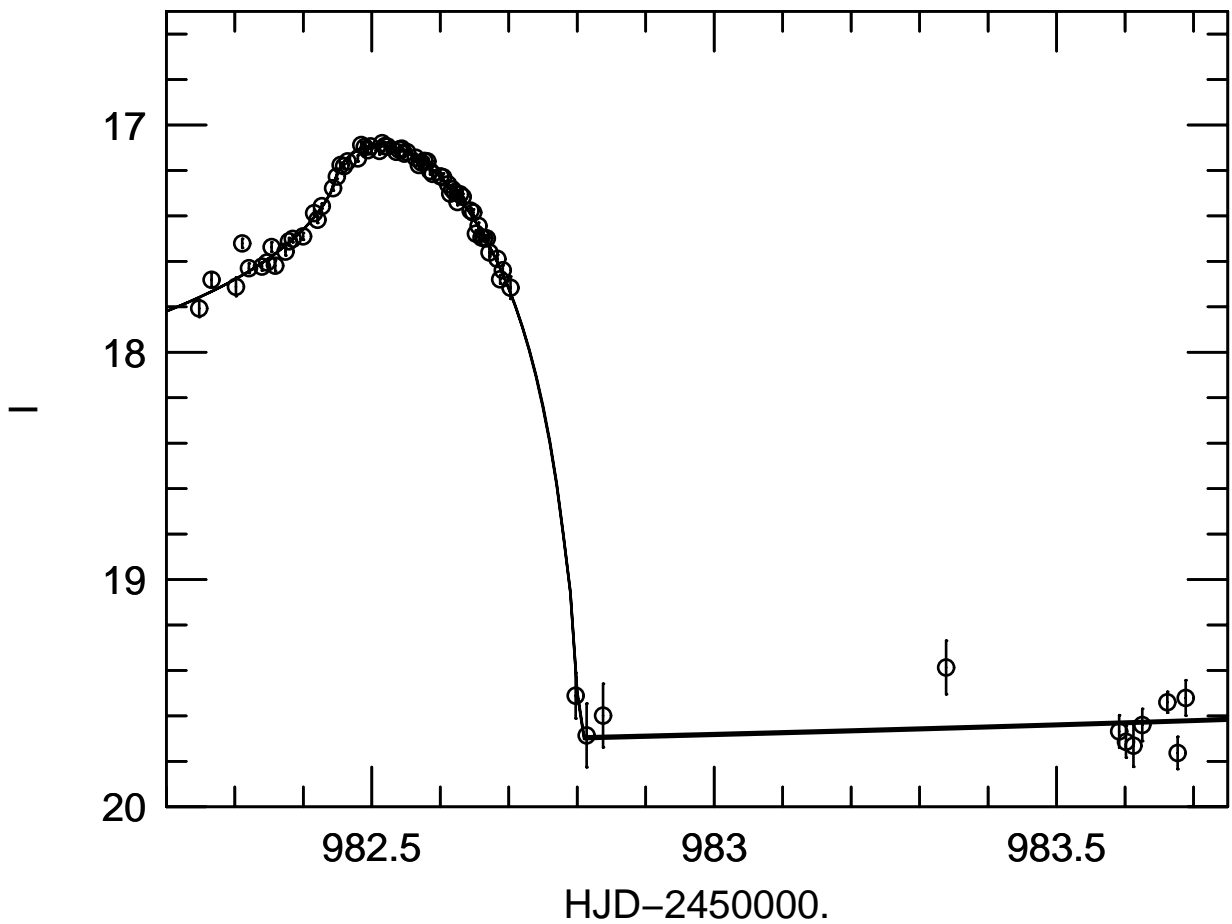
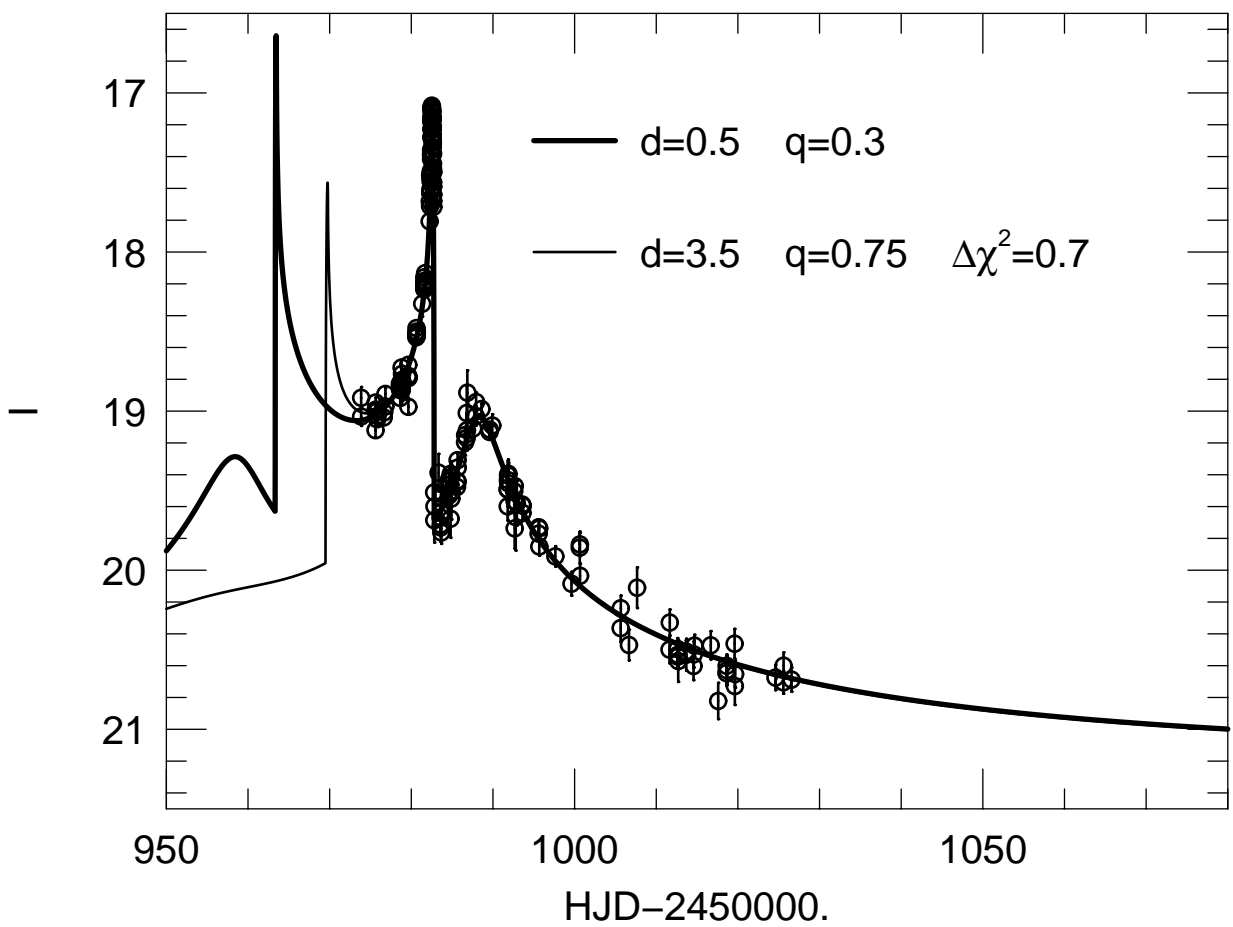
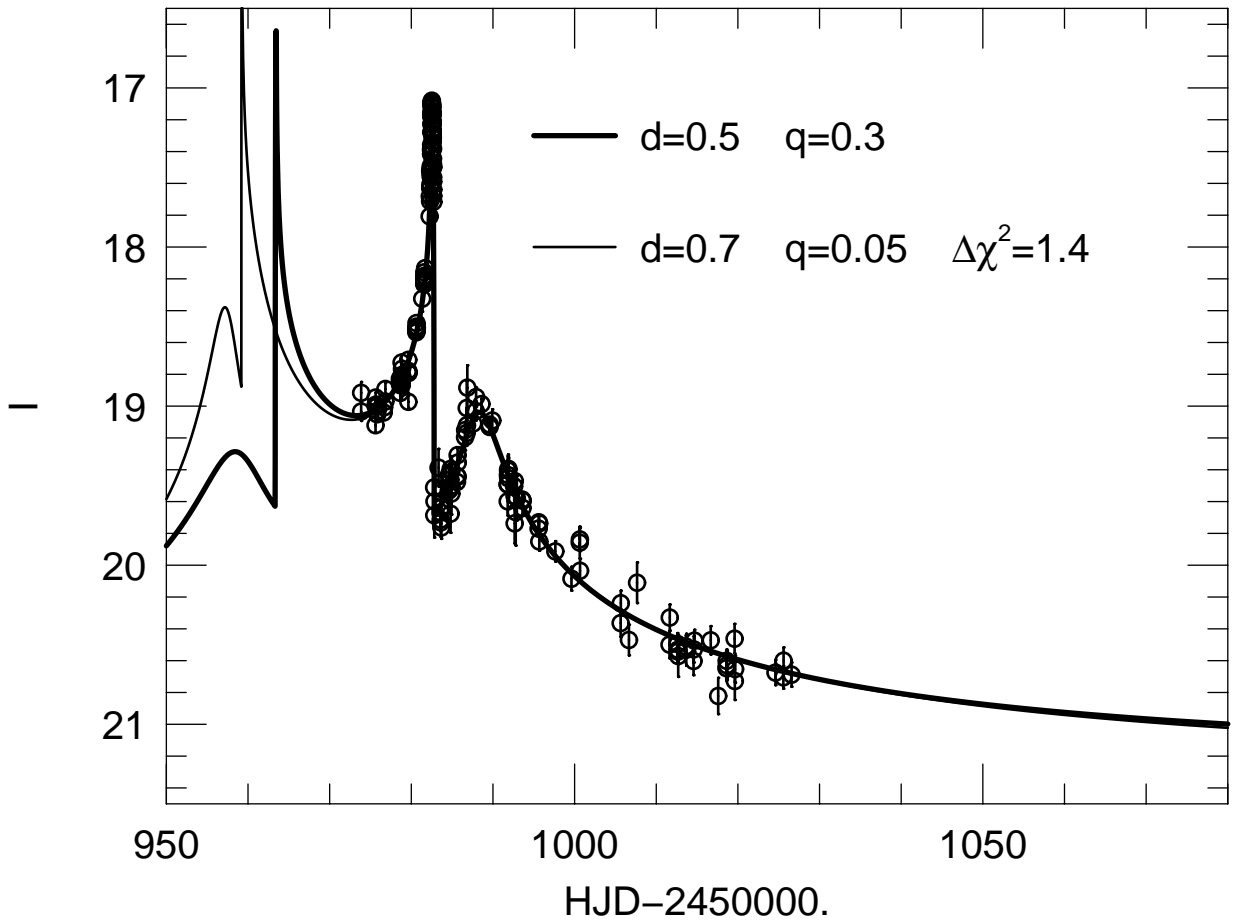
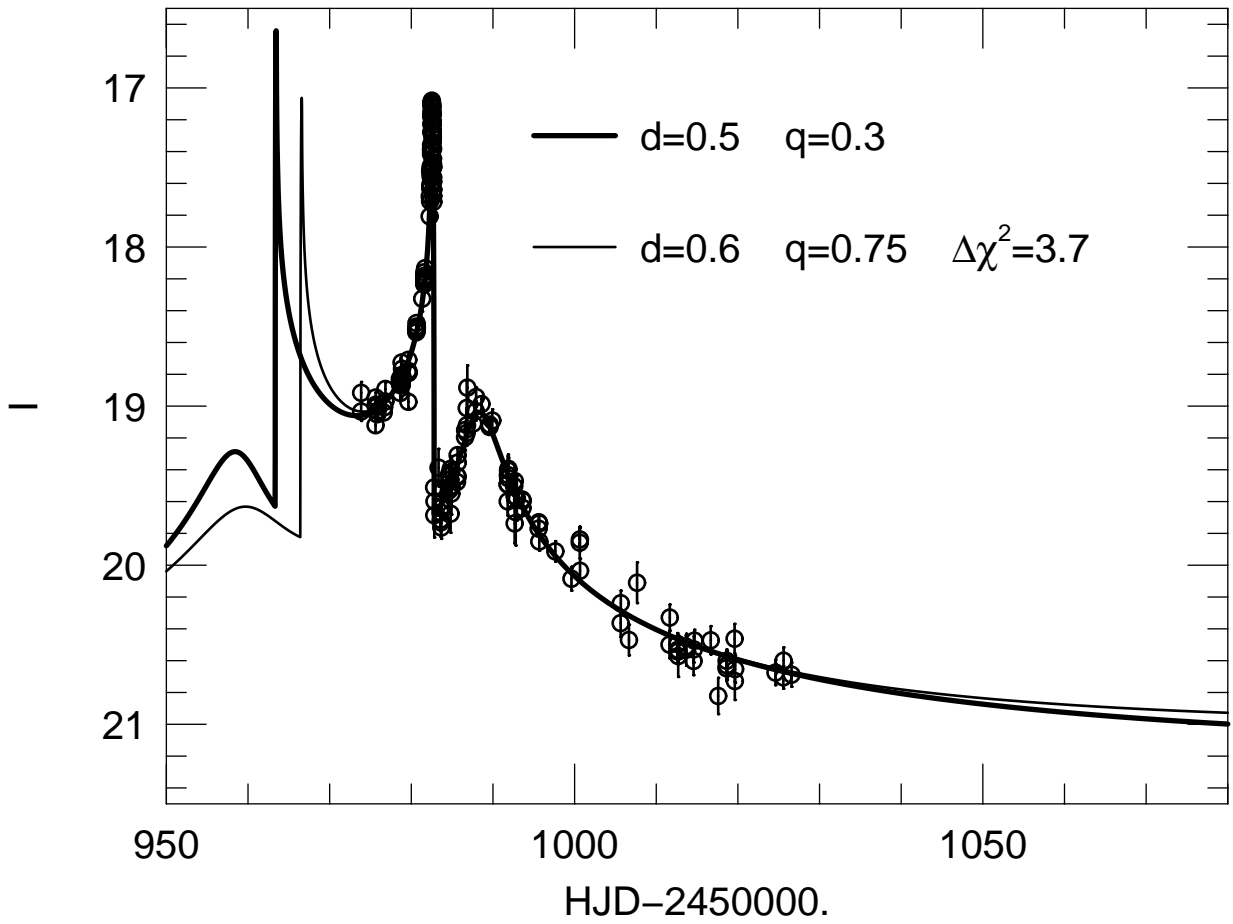
\end{document}